\documentclass[10pt,aps,prd,showpacs,amsmath,amssymb,floatfix]{revtex4}

\usepackage{graphicx}   

\begin{document}

\title{A comparison of methods for gravitational wave burst searches from LIGO and Virgo}

\author{F.~Beauville$^8$, 
M.-A.~Bizouard$^{10}$, 
L.~Blackburn$^3$, 
L.~Bosi$^{11}$, 
L.~Brocco$^{12}$, 
D.~Brown$^{2,7}$,
D.~Buskulic$^8$, 
F.~Cavalier$^{10}$,
S.~Chatterji$^2$, 
N.~Christensen$^1,^{13}$, 
A.-C.~Clapson$^{10}$, 
S.~Fairhurst$^7$, 
D.~Grosjean$^8$, 
G.~Guidi$^9$, 
P.~Hello$^{10}$, 
S.~Heng$^6$,
M.~Hewitson$^6$,
E.~Katsavounidis$^3$, 
S.~Klimenko$^5$,
M.~Knight$^1$, 
A.~Lazzarini$^2$,
N.~Leroy$^{10}$,
F.~Marion$^8$,
J.~Markowitz$^3$, 
C.~Melachrinos$^3$, 
B.~Mours$^8$, 
F.~Ricci$^{12}$, 
A.~Vicer\'e$^9$, 
I. Yakushin$^4$, 
M.~Zanolin$^3$ \\[.3cm]
\begin{center}The joint LSC/Virgo working group\end{center}}
\vspace{.3cm}
\affiliation{$^1$ Carleton College, Northfield MN 55057, USA}
\affiliation{$^2$ LIGO-California Institute of Technology, Pasadena CA 91125, USA}
\affiliation{$^3$ LIGO-Massachusetts Institute of Technology, Cambridge MA 02139, USA}
\affiliation{$^4$ LIGO Livingston Observatory, Livingston LA 70754, USA}  
\affiliation{$^5$ University of Florida - Gainesville FL 32611, USA}
\affiliation{$^6$ University of Glasgow, Glasgow, G12~8QQ, United Kingdom} 
\affiliation{$^7$ University of Wisconsin - Milwaukee, Milwaukee WI 53201, USA}
\affiliation{$^8$ Laboratoire d'Annecy-le-Vieux de Physique des Particules, Chemin de Bellevue, BP 110, 74941 Annecy-le-Vieux Cedex, France }
\affiliation{$^9$ INFN Sezione Firenze/Urbino Via G.Sansone 1, I-50019 Sesto Fiorentino; and/or Universit\`a di Firenze, Largo E.Fermi 2, I-50125 Firenze and/or Universit\`a di Urbino, Via S.Chiara 27, I-61029 Urbino, Italia}
\affiliation{$^{10}$ Laboratoire de l'Acc\'el\'erateur Lin\'eaire, IN2P3/CNRS-Universit\'e de Paris XI, BP 34, 91898 Orsay Cedex, France}
\affiliation{$^{11}$ INFN Sezione di Perugia and/or  Universit\`a di Perugia, Via A. Pascoli, I-06123 Perugia, Italia}
\affiliation{$^{12}$ INFN Sezione di Roma  and/or Universit\`a ``La Sapienza",  P.le A. Moro 2, I-00185 Roma, Italia}
\affiliation{$^{13}$ European Gravitational Observatory (EGO), Via E. Amaldi, I-56021 Cascina, Italia}


\today


\begin{abstract}
The search procedure for burst gravitational waves has 
been studied using 24 hours of simulated data in a network 
of three interferometers (Hanford 4-km, Livingston 4-km and 
Virgo 3-km are the example interferometers).
Several methods to detect burst events developed in the LIGO Scientific Collaboration 
(LSC) and Virgo collaboration have been studied and compared.
We have performed coincidence analysis of the triggers 
obtained in the different interferometers with and without 
simulated signals added to the data.
The benefits of having 
multiple interferometers of similar sensitivity are demonstrated by
comparing the detection performance of the joint coincidence 
analysis with LSC and Virgo only burst searches.
Adding Virgo to the LIGO detector network can increase by 50\%
the detection efficiency for this search.
Another advantage of a joint LIGO-Virgo network is the ability 
to reconstruct the source sky position. The reconstruction accuracy 
depends on the timing measurement accuracy of the events in each 
interferometer, and is displayed in this paper with a fixed source position example.
\end{abstract}

\maketitle

\section{Introduction}
The recent progress in the commissioning of gravitational wave
interferometric detectors raises the possibility for performing joint
data analysis between different collaborations.
Indeed, the LIGO (two 4-km and one 2-km) \cite{ref:LIGO} interferometers are at their design sensitivity, while the Virgo (3-km) interferometer 
sensitivity is approaching its design expectations \cite{ref:Virgo}. 
Moreover, efforts are currently being conducted in order to increase the duty cycle of 
each interferometer, and this will provide extended joint data taking periods.

Gravitational wave (GW) burst events are characterized by their short duration (a few ms)
and the absence of reliable theoretical predictions for their waveforms.
The sources of GW burst events are numerous: 
massive star core collapse \cite{ref:zwerger} \cite{ref:dimmelmeier02} \cite{ref:ott04} \cite{ref:shibata}, 
the merging phase of coalescing 
compact binary systems forming a single black hole \cite{ref:flanagan} \cite{ref:baker} 
\cite{ref:pretorius} \cite{ref:campanelli}, black hole ring-down \cite{ref:kokkotas},
astrophysical engines that generate gamma-ray burst \cite{ref:meszaros}, and even 
neutron star oscillation modes and instabilities. 
Newly born rapidly rotating neutron stars may emit GWs at the frequencies of the
quasi-normal modes during the star's first minutes, and are expected to emit damped sinusoidal 
waveforms \cite{ref:ferrari}. 
More exotic sources are possible, such as GW bursts emitted by cosmic strings \cite{ref:damour}.
Except for a few sources, the burst waveforms are poorly known and their 
amplitude is rather low.
In this study, we did not consider well-modeled waveforms whose search is optimally done using matched filtering techniques. 
We concentrate on sources emitting poorly known waveforms.

It has been long acknowledged \cite{ref:coh89} 
\cite{ref:arnaud03_1} that 
GW burst detection can be enhanced by using 
information contained in a network of detectors spread around the 
world. Due to the time shortness of the burst GW and their weakness
we are obliged to work with quite high false alarm rates in each 
interferometer. With such a configuration, it is necessary to use a 
network of interferometers to reduce the false alarm rate (FAR).
This is also the only way to disentangle a real GW burst event from 
transient noise events with sufficient confidence.
Several methods and strategies can be used. 
There exist two categories of network data analysis: coincidence or 
coherent filtering. 
In the first approach, each interferometer data stream is analyzed,
providing a list of event triggers. Triggers from the different
interferometers are then compared to check if they are consistent with 
a real GW burst source. 
A coherent analysis uses all interferometer information by 
combining the input data streams or the filtered data streams into
one single stream. 
The coincidence analysis is simpler to apply than the coherent 
analysis since it does not require the exchange of the detector data 
streams; this technique has been commonly used \cite{ref:igec} \cite{ref:BLS2}.
On the other hand, the coherent analysis is expected to be, a priori, 
more powerful than the coincidence approach since a depreciated signal in a
single interferometer can contribute to the global network output.
In this investigation, we focused on the coincidence analysis, trying to study all different aspects (including source location) since it is the simplest method to initially apply to multi-detector data. 
A similar comparison study for a coherent search is also underway for the LIGO-Virgo network.

The LSC burst search pipeline already performs coincidence analysis
using LIGO's two well aligned 4-km interferometers, and its 2-km interferometer. 
In addition, several joint searches with other detectors using real data 
have been done or are underway;
an initial joint LIGO-TAMA coincidence search using real data has been reported
in \cite{ref:BLT}.
Joint searches for LIGO-GEO and LIGO-AURIGA are underway. 
Similar efforts are carried out for binary inspiral GW searches. 
In the study presented in this paper, we analyze the potential benefits 
of a combined LIGO-Virgo detector network burst search; our study here
is conducted using simulated data. This network is composed of
three detectors of similar sensitivity. In actuality, having 
similar sensitivity plays a crucial role in the performance 
of a network (that is the reason why we did not consider the LIGO Hanford 2-km interferometer
in this study. In addition, it could not be used to eliminate fake events due to environmental noise sources 
or detector artifacts since this study is carried out with simulated data and not real data). 
Nevertheless, while the LIGO detectors have the same orientation being separated by only 3000 km,
the Virgo detector orientation and its distance with respect to the LIGO detectors imply 
that Virgo and the LIGO detectors do not have a maximal sensitivity simultaneously for the
same region of the sky. 
Actually they are somewhat ``orthogonal''. 
It is then important to determine 
the gain of a coincidence search in such a network with respect to a search conducted using only the 
LIGO detectors.
Note that a similar comparison study for the inspiral GW search has been carried out \cite{ref:inspiral}.

The paper is organized as follows. Section \ref{sec:section2} contains a description of
the characteristics of the data used for this analysis, including both noise and the
source waveforms; this is the data that has been used for studying the different 
pipelines and the coincidence analysis.
Section \ref{sec:section3} summarizes the performance of seven different burst 
filters considered in this joint data analysis \cite{ref:zanolin04}. 
After having defined a benchmark, we evaluate and compare the intrinsic 
performance of each filter, and characterize each pipeline as it is applied on both the LIGO 
and Virgo data streams.
We address, at the end of this section, the issue of the burst signals' 
parameter space coverage by the different pipelines. We show, using simulated
burst waveforms of different types, that the description of a burst GW with few statistical
parameters (as has been proposed in the literature) is not complete.
We have investigated several strategies to carry out a coincidence analysis, 
taking into account that the three detectors do not have the same sensitivity with respect to sky
location for a particular source. The requirement for detecting a
source in all the interferometers greatly decreases the FAR, 
but the detection efficiency may also drop drastically if
the source is located in a region of the sky for which at least one 
detector has poor antenna response.
We find that requiring detection by any two of the three detectors gives improved 
detection efficiency compared to triple-coincidence detection at the same false alarm rate.
Section \ref{sec:section4} gives a summary of what can be achieved in such a 
network applying the different filters used by the two detector teams.
We consider a source located at a fixed sky location (in the direction of the galactic center).
We show the performance gain when adding a third interferometer of 
similar sensitivity but not well aligned with the first two. 
We have also estimated the gain of adding the event frequency 
information in the coincidence analysis. 
The benefits and disadvantages of a logical combination of filtered 
output obtained for each of the detector data streams (the so called ``AND'' 
and ``OR'' analysis) have been investigated too and are summarized 
in Section \ref{sec:section5}.
Working with three interferometers spread over the world allows one to
estimate the source sky location. In this study, we apply a method
developed in Virgo \cite{ref:VSL} for estimating the source location. 
As an elementary building block of an all-time, all-sky search we decided 
to look for a source that emits from a fixed location in the sky.
The burst ``repeater'' source was placed at the 
center of the Galaxy and the performance modulation due to Earth's rotation 
has been studied with 24 hours of simulated data. These results are given in 
Section \ref{sec:section6}. Our results are summarized in Section \ref{sec:section7}.

\section{Data set}
\label{sec:section2}
\subsection{Interferometer noise modeling}
\begin{figure}[h]
\begin{center}
\includegraphics[width=12cm]{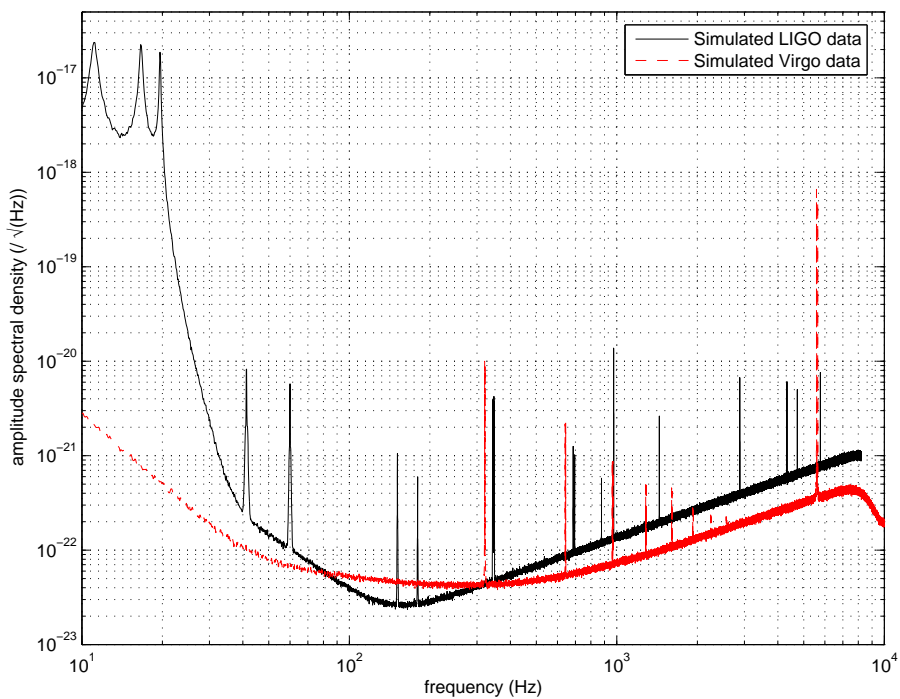} 
\caption{\label{fig:H1Noise} LIGO (solid line) and Virgo (dashed line) simulated noise spectrum used 
in this study. The distortion close to the Nyquist frequency in the Virgo spectrum is due 
to the use of a low pass filter applied before down-sampling the Virgo data generated at 
40 kHz down to 20 kHz.}
\end{center}
\end{figure}

Simulated noise data streams have been generated for the LIGO Hanford (H1), 
LIGO Livingston (L1) and Virgo (V1) interferometers independently using LSC \cite{ref:LAL}
and Virgo \cite{ref:siesta} simulation tools. 
In both cases, a Gaussian, stationary and colored noise model has been used 
such as to reproduce the design sensitivity curve of the LIGO and Virgo 
interferometers as shown in FIG. \ref{fig:H1Noise}. 
This includes thin thermal resonances (wire violin modes and mirror thermal noise)
as foreseen in each detector (thermal noise modeling has been used for the Virgo noise 
generation while a random phase modulation technique has been used for the LIGO 
noise simulation). The LIGO noise also includes lines due to electrical power supply. 
In Virgo the 50Hz harmonics lines are subtracted in the main 
reconstructed gravitational wave strain applying a deterministic algorithm using 
an auxiliary channel which monitors the power supply of Virgo \cite{ref:flaminio}.
Two data sets have been produced. Initially 3 hours of H1 and V1 noise 
were created and used to characterize the burst filters and cross-check the consistency of the 
different pipelines used in the LSC and Virgo. Then, 24 hours of noise data have been 
produced for H1, L1 and V1 interferometers.
These 24-hour long data streams have been used for the coincidence analysis presented in
this paper.  

\subsection{Signals}

\subsubsection{Waveforms and parameter space}
\label{sec:parameter}
Because many gravitational wave burst sources are currently not modeled in sufficient 
detail to allow matched filtering, most gravitational wave burst detection algorithms are 
designed to be insensitive to the details of the waveforms, and search through broad 
time and frequency ranges.
However, in order to carry out a comparison of the different filters, we decided 
to use in this study three statistical variables as proposed in \cite{ref:cheese} to 
parametrize a burst signal. 
These three parameters are the time duration ($\Delta \tau$), the frequency bandwidth ($\Delta \phi$) 
and the peak frequency ($\phi$) of the signal as defined using the first and second 
moments of the signal energy distribution in the time and frequency domains. 
Actually, the definition of the time duration and frequency bandwidth used in this paper 
differs from the one one proposed in \cite{ref:cheese}: 
they are defined by the interval containing 50\% of the signal energy, dropping 25\% of the 
signal energy on both sides of the distribution.
Note that $\Delta \phi$ and $\Delta \tau$ are constrained by some uncertainty relation \cite{ref:cheese} 
\cite{ref:flandrin}:
\begin{equation}
\label{eq:heisenberg}
\Delta \tau  \Delta \phi > \frac{\Gamma}{4\pi} ~~~~~~~~ \Gamma = 0.59...
\end{equation}
This equation (also known as Heisenberg-Gabor uncertainty relation) indicates 
that a signal cannot have arbitrarily small duration and bandwidth simultaneously.\\

The range values of those three quantities define 
a parameter space inside which the burst filters have to be as efficient as possible. 
The boundaries have been chosen taking into account theoretical predictions available 
in the literature and by the noise spectra shown in FIG. \ref{fig:H1Noise}. 
They are given in TABLE \ref{tab:parameter_space}.

\begin{table}[h]
\begin{tabular}{c|c|c|c}
\hline
         & duration ($\Delta \tau$) & peak frequency ($\phi$) & frequency bandwidth ($\Delta \phi$)\\
\hline
minimum value    &  0.5 ms       &  50 Hz         & 5 Hz \\
maximum value    &  50  ms       &  2000 Hz       & 1000 Hz \\
\hline
\end{tabular}

\caption{\label{tab:parameter_space}Boundaries of the parameter space for the burst signals
studied. Time duration, peak frequency and bandwidth are the three parameters 
that have been chosen to describe the main features of expected burst signal waveform.}
\end{table}

\begin{figure}[h]
\begin{center}
\includegraphics[width=12cm]{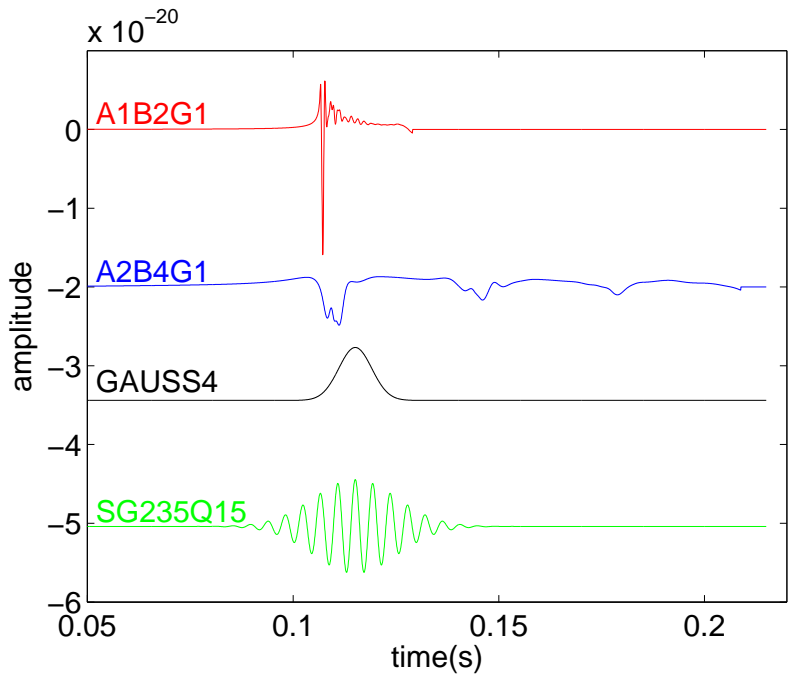}
\caption{\label{fig:injections}Waveform families of burst sources used 
in this study to estimate the performance of the different pipelines. DFM A1B2G1 and DFM A2B4G1: gravitational wave 
emitted during core collapse simulated by Dimmelmeier, Font and Mueller \cite{ref:dimmelmeier02}. 
GAUSS4: Gaussian peak. SG235Q15: Sine-Gaussian signal.}
\end{center}
\end{figure}

\begin{figure}[h]
\begin{center}
\includegraphics[width=12cm]{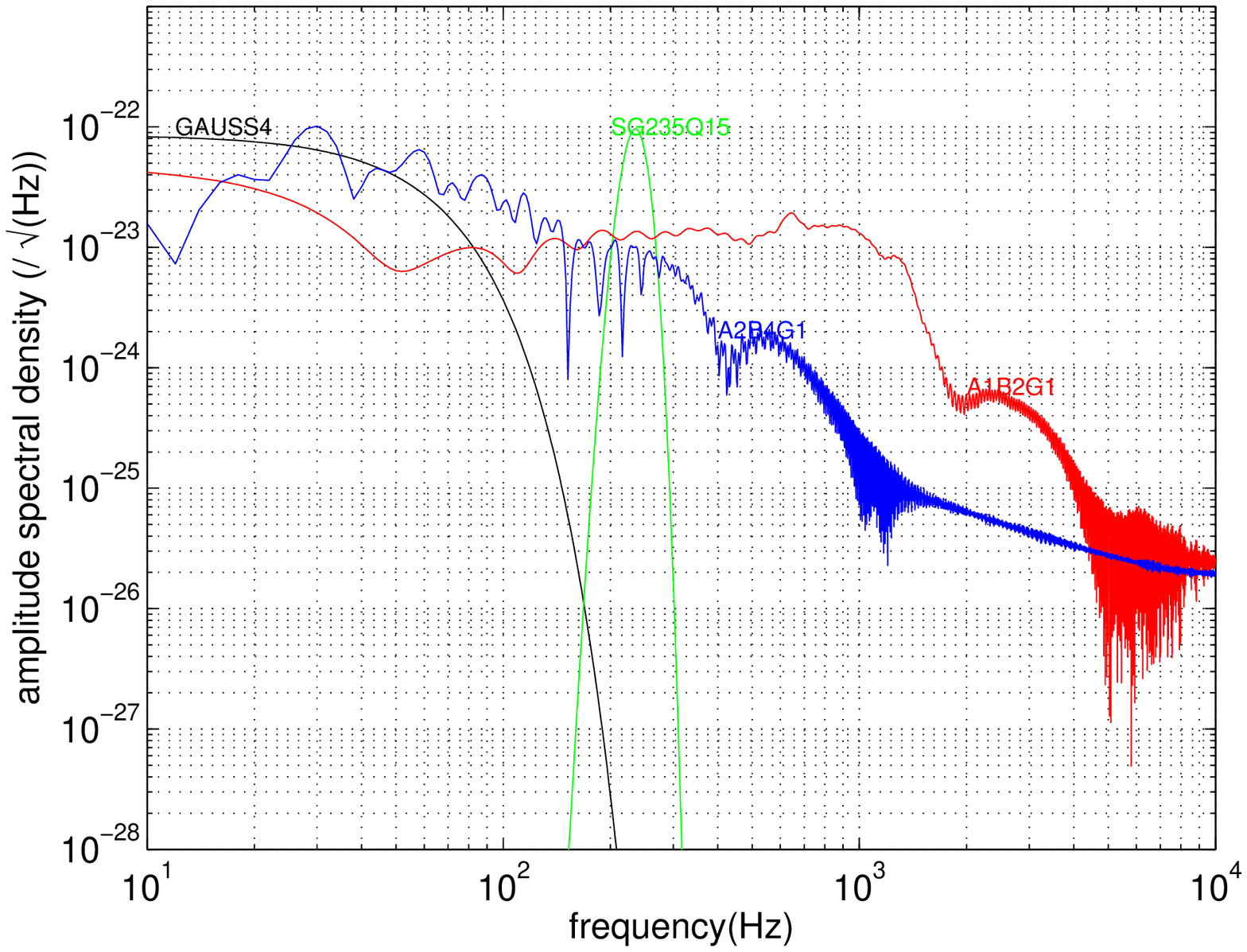}
\caption{\label{fig:spectra}Spectrum of the waveform families of burst sources used 
in this study to estimate the performance of the different pipelines. DFM A1B2G1 and DFM A2B4G1: gravitational wave 
emitted during core collapse simulated by Dimmelmeier, Font and Mueller \cite{ref:dimmelmeier02}. 
SG235Q15: Sine-Gaussian signal. GAUSS4: Gaussian peak.  All waveforms have been normalized such that $h_{rss} = 3.59 \times 10^{-22}  \sqrt{Hz}$.
This value corresponds to a SN source located at 10 kpc emitting a A1B2G1 waveform.}
\end{center}
\end{figure}

In order to fully characterize our burst filters it is necessary to evaluate their performance 
using just a few waveforms that have different features and are well spread over the parameter space. 
We have used three different families of waveforms. First, broad-band Gaussian signals are of 
interest because many predicted core collapse simulation waveforms do have large peak 
structures corresponding to the bounce. 
Sine-Gaussian waveforms are narrow-band signals that allow one to test a given 
frequency region. Finally, more complex waveforms predicted by core collapse simulation, 
as described in \cite{ref:dimmelmeier02}, have been used; we have chosen two of them 
obtained with different assumptions on the model parameters. In both cases the equation 
of state of the ideal gas is stiff, but in one case the differential rotation is small 
leading to type I (regular collapse) waveform (large negative peak followed by a ring down phase). 
In the second case, the initial state is characterized by a rapidly rotating star with a
large differential rotation producing a type II waveform characterized by multiple bounce collapse. 
The 8 chosen waveforms, whose parameters are given in TABLE \ref{tab:injection_parameters}, 
are represented in FIG. \ref{fig:injections}. 
Their frequency spectra are shown in FIG. \ref{fig:spectra}.

\begin{table}[h]
\begin{tabular}{l|c|c|c}
\hline
Name     & time domain formula      & parameter value  & used for coincidence analysis\\
\hline
GAUSS1 &    $h(t) = \alpha~ exp(\frac{-(t-t_0)^2}{2 \sigma ^2} )$  & $\sigma = 1$ ms & yes\\
GAUSS4 &                                                   & $\sigma = 4$ ms & yes\\
\hline
SG235Q5  &  $h(t) = \alpha~ exp(\frac{-2(t-t_0)^2 \pi^2 f^2}{Q ^2} ) cos (2\pi f (t-t_0))$ &  f=235 Hz, Q=5 & no\\
SG235Q15 &           & f=235 Hz, Q=15 & yes\\
SG820Q5  &           & f=820 Hz, Q=5 & no\\
SG820Q15 &           & f=820 Hz, Q=15 & yes\\
\hline
DFM A1B2G1   &           & A=1, B=2, G=1 & yes\\
DFM A2B4G1   &           & A=2, B=4, G=1 & yes\\
\hline
\end{tabular}

\caption{\label{tab:injection_parameters} Definition and parameter values of the 8 
waveforms used to estimate the burst filters performance. For the DFM A1B2G1 and DFM A2B4G1 signals
the A, B and G parameters are defined in \cite{ref:dimmelmeier02}. 
Information in the far-right column indicates whether the signal was used in the coincidence analysis described in Section \ref{sec:section4}.}
\end{table}

The normalization of the signals can be done either by fixing the intrinsic signal strength
($h_{rss}$) or by the intrinsic signal to noise ratio (SNR) of the signal seen in a detector. 
The latter quantity is more suitable if one wants to compare a filter's performance in interferometers 
with different sensitivity. 
On the other hand, in the case of a network analysis, it is mandatory to consider a source 
of a given intrinsic amplitude. This quantity is defined
by:
\begin{equation}
h_{rss} = \sqrt {\int_{-\infty}^{\infty} |h(t)|^2 dt}
\label{eq:hrss}
\end{equation}

The intrinsic SNR seen in a detector is given by the SNR of a matched filter: 
\begin{equation}
\rho = 2 \sqrt { \int_0^{\infty} \frac{|h(f)|^2}{S_h(f)} df}
\label{eq:SNR}
\end{equation}

where $S_h(f)$\, is the one-sided power spectrum of the detector noise.\\

For the estimation of the pipelines' performance given in Section \ref{sec:section3} 
we have considered that the detectors are optimally oriented with respect to the source. 
The free parameters of the different pipelines studied here have been chosen
such as to maximize the detection efficiency of the eight chosen waveforms. 
In Section \ref{robustness} we address the issue of a possible dependence 
of the pipeline parameters' tuning upon the waveforms considered. 
For that purpose, we have extended the set of waveforms, and studied a larger set of core 
collapse simulated waveforms (50). 
We have also added 50 waveforms 
which fill the region of the parameter space outside the ``minimal uncertainty'' region.
They are generated using band passed white noise. 

\subsubsection{Coincidence analysis injections}
\label{sec:coininj}
\begin{figure}[h]
\begin{center}
\includegraphics[width=10cm]{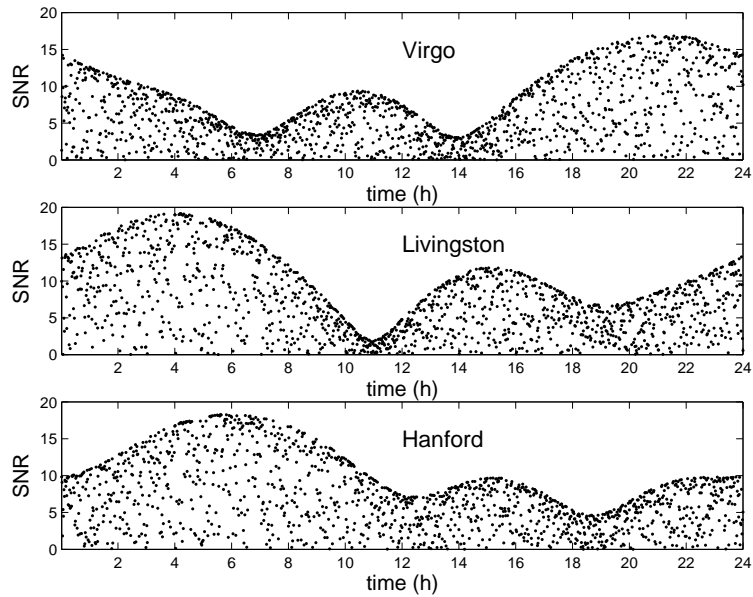}
\caption{\label{fig:BPmodu}Signal to noise ratios of DFM A1B2G1 signals emitted by a source
located in the direction of the galactic center over 24 hours. The SNRs were computed for the H1,
L1 and V1 detectors, taking into account their antenna pattern at the signal 
arrival time (note the envelope of the distributions). In this study, the source is polarized with
polarization angle uniformly distributed; this produces the distribution of values inside the envelope.} 
\end{center}
\end{figure}

For the coincidence analysis described in the Sections \ref{sec:section4}, \ref{sec:section5} and \ref{sec:section6}
we located the source in the direction 
of the galactic center. The waveforms are linearly polarized, with polarization angle 
uniformly distributed. In addition to the unknown polarization angle, the response of 
the detector to a GW depends on the sky position of the source. A source of constant
intrinsic emission strength located in the direction of the galactic center would emit signals whose amplitude 
measured by a detector would then vary as shown in FIG. \ref{fig:BPmodu}. We have used 24 hours of 
simulated data in order to study, on average, the effect of the earth rotation on the detection
performance.
Due to the use of non-astrophysical waveforms (with the exception of supernova core collapse) 
we cannot use the distance to fix the strength of the signal. 
We have then normalized the 6 waveforms such that over 24 hours there is only one injected event 
seen with a SNR of 10 or greater in all the three detectors. For the supernova core collapse, this 
scaling corresponds to a distance of 4.8 kpc for DFM A1B2G1 and 3.6 kpc for DFM A2B4G1.

\section{Burst search methods and their performance}
\label{sec:section3}

In this investigation, we have studied different burst detection algorithms. We compared their performance 
in order to determine if they cover the variety of possible astrophysical waveforms.
Indeed, before performing
a network analysis with some of them it is important to know how they behave individually. 
For each filter, we measure its detection efficiency as a function of the false alarm rate.
Then, the arrival time of the signal in each interferometer has to be determined with very 
good accuracy (of the order of 1 ms) in order to be able to reconstruct the position of 
the source in the sky. 
In the following, we briefly describe the different pipelines used in this study.
It is important to note that some of these filters require
whitened data. This pre-processing stage is performed both for the LSC and Virgo pipelines using 
a linear predictive error filter. These have been developed in each collaboration, see for instance
 \cite{ref:cuoco} \cite{ref:QT}

\subsection{Pipeline description}
We utilized search methods operating in the time and time-frequency domain. 
In addition, correlator filters were also applied. 
Two of these pipelines (KW and QT) have been developed in the LSC while 
the others have been developed in Virgo.

\subsubsection{Excess power methods}
\begin{itemize}

\item{Power Filter (PF)}\\
PF \cite{ref:PF} searches on whitened data for a power excess using different time analysis 
windows and different frequency bands chosen such that the product of the frequency width and the 
time duration remains constant. In order to cover the parameter space 5 time windows have 
been chosen: 64, 128, 256, 512 and 1024 bins for LIGO and Virgo data; the bin size is defined as the 
sampling time for the data, which is $1/16384$ s for LIGO and $1/20000$ s for Virgo. The windows duration is 
hence a little different for the two data sets because of the different data sampling rates.
PF has been tuned, in this study, to cover
all frequencies between 40 Hz up to 2000 Hz.
In the time-frequency plane the PF statistics are derived by taking the logarithm of the spectrogram.

\item{Kleine Welle (KW)}\\
KW searches for excess signal energy in the dyadic wavelet
decomposition of a whitened time series. The wavelet transformation is
evaluated at distinct scales (which is directly related to frequency) 
and involves stretching and contracting a mother wavelet (in
this case the {\em Haar} wavelet) by the scale factor before using it to
filter the time series \cite{ref:KW}. In this study, the frequency range covered 
by KW was from 64 Hz to 1064 Hz for H1 and L1, and from 78 Hz up to 1250 Hz for V1 data.
As the input time series is whitened by use of a
linear predictive filter prior to the wavelet decomposition, the wavelet
coefficients calculated at each scale follow a Gaussian distribution and
are normalized to unity variance. KW identifies events in the
time series by clustering large-amplitude coefficients which are nearby in
time and scale.

\end{itemize}

\subsubsection{Time domain filters}
\begin{itemize}
\item{Mean Filter (MF)}\\
MF searches for an excess in a moving average computed on whitened data. 
Ten different analysis windows, 
with durations varying from 0.5 ms up to 10 ms, are used. The moving stride of each window is 1 bin.  
It has been shown that the analysis window
size has to be matched on the duration of the signal to have the best detection result \cite{ref:arnaud03_1}. 
Note that MF behaves like a low pass filter and, given the choice of the windows' size, 
MF could not explore the frequency domain beyond 1 kHz. A threshold 
is applied on each analysis window output, producing triggers which may overlap. 
They are then clustered if they coincide within a time window equals to 12 times the size 
of the analysis window. This value corresponds to the optimal signal 
efficiency for the chosen threshold (SNR$>$4) obtained on simulated Gaussian data in which signals, 
such as the ones used in this study, have been injected.
Finally, note that no information about signal frequency is provided.

\item{Alternative Linear Filter (ALF)}\\
The ALF filter aims at detecting large slopes associated with peaks in the data 
in 10 moving analysis windows whose duration 
varies between 0.5 ms and 15 ms. The moving stride of each window is 1 bin. 
The frequency range that ALF can explore is below 1 kHz.
A straight line is fit to the data in an analysis window. 
The 2 non-independent fit results, namely the line slope and the offset value, are then 
combined quadratically to define the ALF statistics. Concerning the optimal analysis window 
size to detect a peak, it is has been shown that the window size has to be 30\% larger than the duration
of the peak. This effect is due to the noise standard deviation which is reduced
if the computation is done over a larger window \cite{ref:pradier01}. 
The pipeline of ALF is identical to that of the MF filter concerning the events clusterization.

\end{itemize}

\subsubsection{Correlator filters}
\begin{itemize}
\item{Peak Correlator (PC)}\\
This is a matched filter using Gaussian waveform templates \cite{ref:arnaud99}. 
The Gaussian templates have been
chosen such that the minimum match is better than 99\% for Gaussian peaks whose $\sigma$\, is 
between 0.2 ms and 6 ms (8 templates). That corresponds to a maximum frequency of about 900 Hz.
Note that PC has been implemented using the Welch overlapping scheme
\cite{ref:welch} which optimizes the computing cost in the case of small template size. The 
events obtained after thresholding are then clustered using only time coincidence
information. 

\item{Exponential Gaussian Correlator (EGC)}\\
The complex Exponential Gaussian Correlator produces a time-frequency
representation of the data by
applying the correlation relation for a list of templates of the same
family:
\begin{equation}
\Phi_{f_0,Q_0}(t) = e^{-2 (\pi f_0/Q_0)^2 t^2} e^{i 2 \pi f_0 t}
\end{equation}
\noindent with $(f_0,Q_0)$ the central frequency and quality factor of the
template. \\

Template selection relies on a tiling algorithm valid for any two-parameter 
matched filter bank \cite{ref:arnaud03_3}.
Two tiling zones have been defined in order to optimize the visibility of non-oscillatory
waveforms while covering a large range of periodic signals with a reduced number of 
templates (122).
The first zone covers $10 \leq f \leq 1000$ Hz and $1 \leq
Q_0 \leq 3$, while the second spans $150 \leq f \leq 1000$ Hz and $1 \leq
Q_0 \leq 17$.
The event duration covered is from 0.2 ms to 50 ms.
Because the frequency coverage is redundant, templates are grouped in frequency bands
50 Hz wide, where they are ordered by decreasing quality factor. 
Events are then identified as local energy excesses. More details can be found in 
\cite{ref:EGC}.

\item{Q-transform (QT)}\\
QT is a multi-resolution search for statistically significant signal 
energy in the time-frequency plane \cite{ref:QT} \cite{ref:KW}.
It is equivalent to a matched filter search for sinusoidal Gaussian
bursts with unknown phase in the whitened data stream. The data is
first whitened by linear predictive filtering. The resulting whitened
data is then projected onto an over-complete basis of Gaussian windowed
complex exponentials within a targeted range of frequency and $Q$
(ratio of frequency to bandwidth). 
For this study, the Q transform was tuned to search the frequency
range from 64 to 1024 Hz and the Q range from 3.3 to 22 corresponding to an event
duration range 0.5 ms up to 50 ms.
Candidate events are then described by the most significant set of non-overlapping
time-frequency tiles that exceed a specified significance threshold.

\end{itemize}

Note that while the QT and EGC filters look similar, they are actually implemented differently. 
The noise normalization is done externally for QT, while EGC performs matched 
filtering. The event extraction is different; EGC includes all neighboring templates
while QT prunes overlapping templates.

\subsection{Event trigger definition}
Each of the burst search filters applies a threshold on the filter output.
The filter output is either defined as an energy in a time-frequency grid or as a SNR 
in the time domain. This is performed for several values of the parameters (frequency band, window size, template). 
For time-frequency methods all nearby pixels are clustered and an energy of the cluster is 
defined as the sum of the energy of the pixels according to a well defined prescription developed 
in each of the methods.
After thresholding, we obtain a collection of events defined by a start time, an end time, 
a peak time, and the corresponding maximum filter output. In addition, filters designed for providing 
localization in the frequency domain (QT, KW and EGC) yield an estimation of the event bandwidth and 
the peak frequency, corresponding to the frequency of the template with the highest SNR.

Almost all the burst filters deal with several analysis windows, templates, or time-frequency 
planes. In the case of a signal, several nearby templates
or time-frequency planes are expected to trigger, corresponding (in actuality) to the same event.
All burst pipelines thus apply an event clustering algorithm that gathers all primary defined
events corresponding to the same signal event. The clustering algorithm usually depends on the
filter type: time-frequency, correlators, and time domain filters. 

\subsection{Detection performance}
\label{sec:performance}
\begin{figure}[h]
\begin{center}
\begin{tabular}{cc}
\includegraphics[width=9cm]{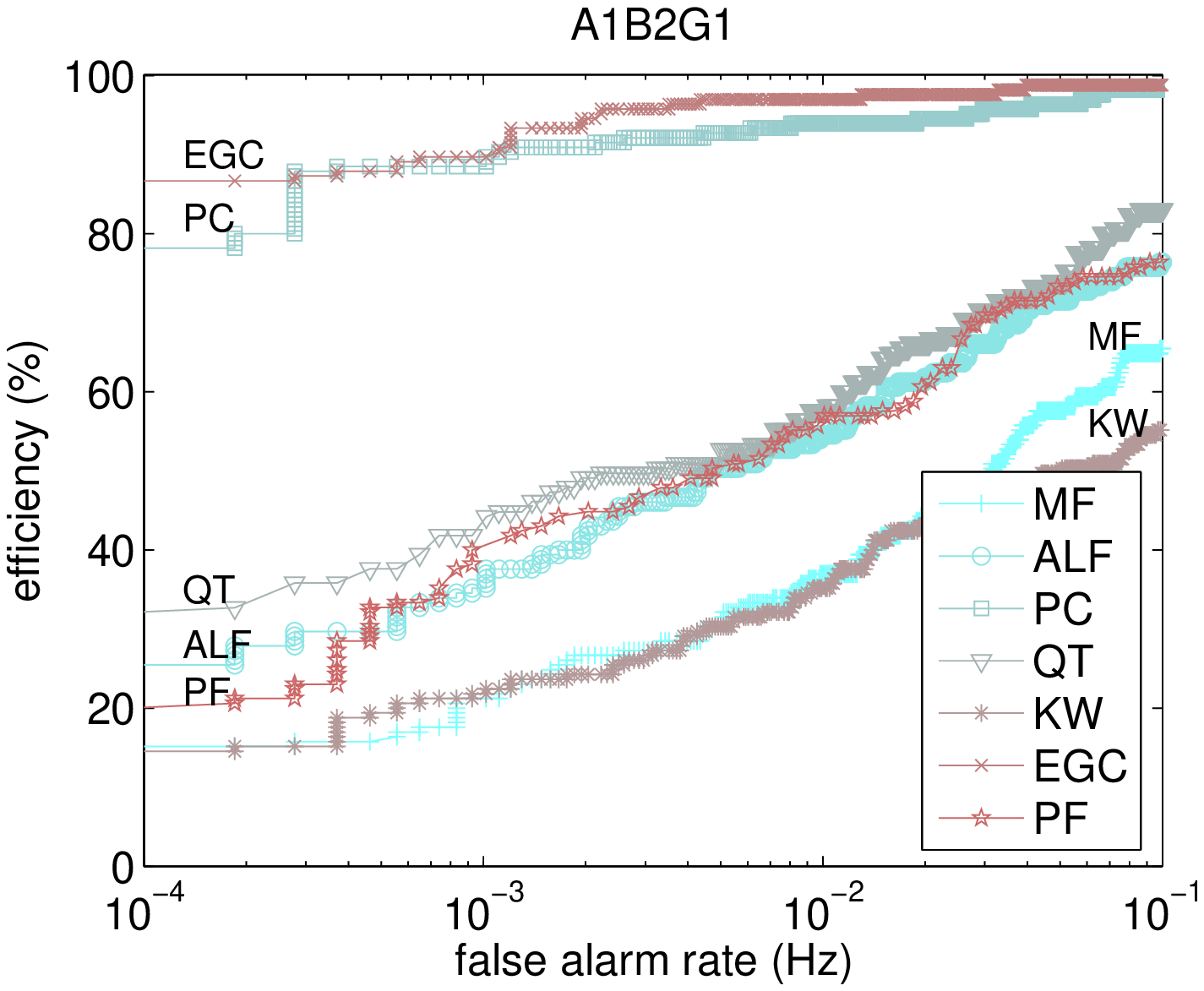} & \includegraphics[width=9cm]{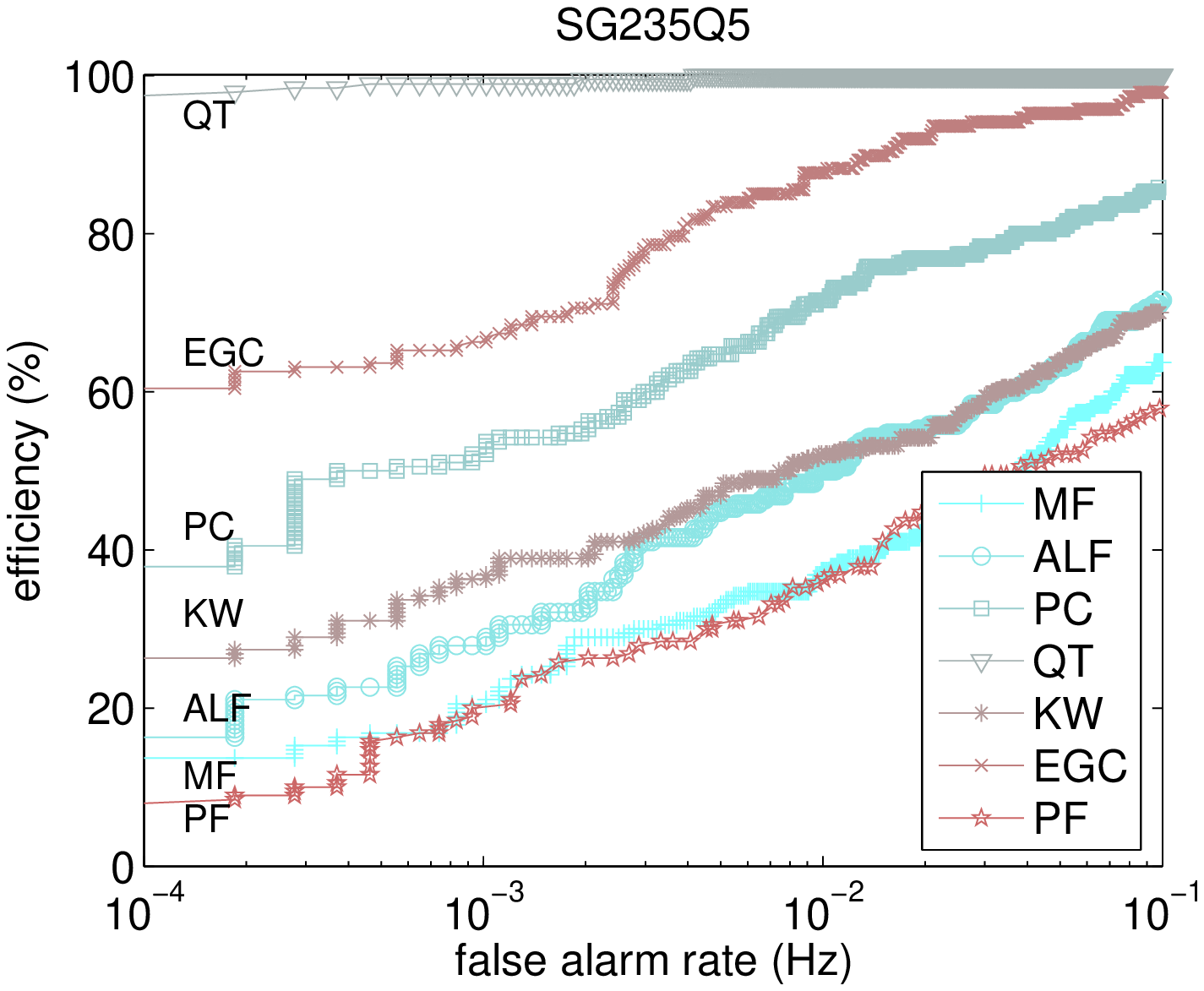}\\ 
\includegraphics[width=9cm]{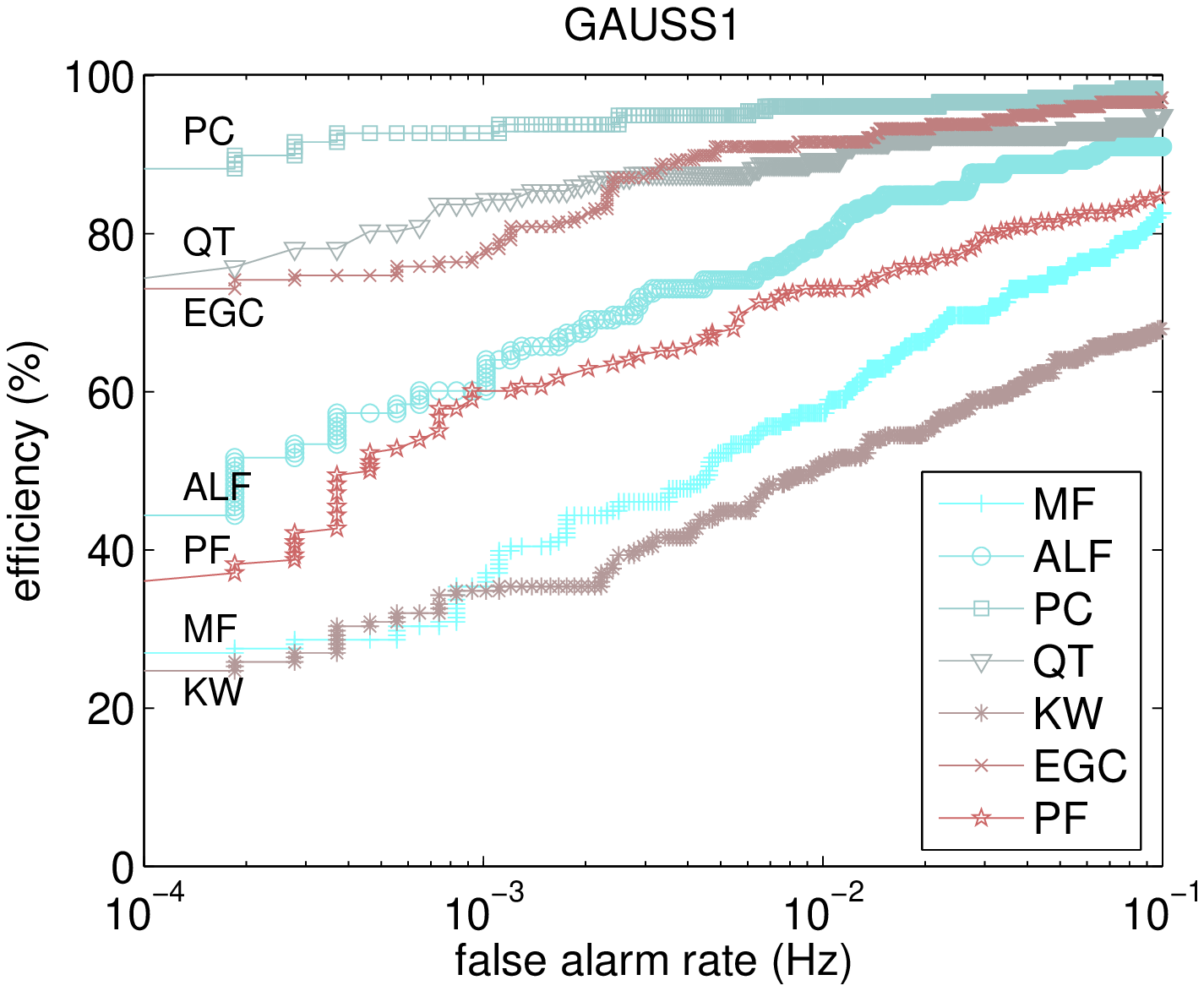} & \includegraphics[width=9cm]{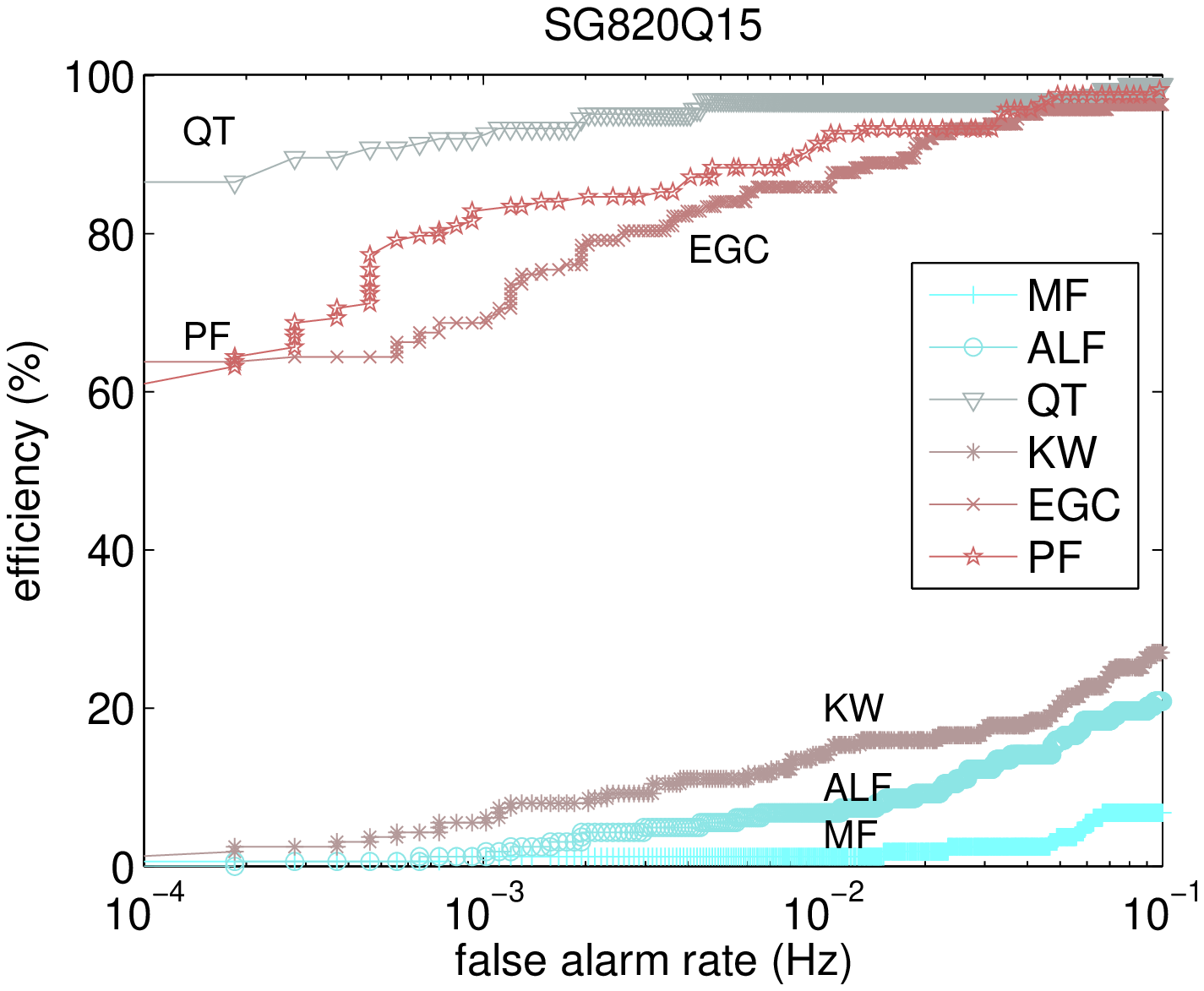}\\ 
\end{tabular}
\caption{\label{fig:rocs} ROC curves obtained from the bank of filters studied 
using 4 different waveforms (DFM A1B2G1, SG235Q5, GAUSS1 and SG820Q15). The 
signal waveforms have been injected into H1 noise with an intrinsic SNR of 7.}
\end{center}
\end{figure}

One way to quantify the detection potential of a burst search filter is 
through the determination of 
Receiver Operator Characteristic (ROC) curves for each filter and for each family of 
waveforms. A ROC curve shows the filter detection efficiency versus the false alarm rate (FAR). 
The FAR is usually a free parameter which is fixed differently with each interferometer data set
according to the type of analysis; a discovery search requires one to have very few false alarms 
in order to 
select golden events which are then studied deeply in detail (waveform, frequency content, arrival time consistency), 
while to better understand the statistical properties of the 
background noise (gaussianity, stationarity) it is necessary have a large FAR. Moreover,
in a network analysis with non-correlated noise in the interferometers, 
a simple coincidence analysis
reduces the final FAR by a few orders of magnitude with respect to the 
single-interferometer FAR (which can be chosen rather high). This is demonstrated in 
Section \ref{sec:section4}. 
The ROC curves for the 8 test-case waveforms have been computed for all 
burst filters and for different values of the strength of the signal. In particular, we checked the 
consistency of the results obtained with H1 and V1 data. In the following we summarize 
the main results about the ROC curves estimated for a FAR varying from $10^{-4}$~ Hz 
(1 false alarm per 3 hours) to $10^{-1}$~ Hz (a single-interferometer FAR 
which can be easily handled in a network coincidence analysis).
FIG. \ref{fig:rocs} shows the ROC curves (using the H1 data) for 4 of the 8 waveforms: 
A1B2G1, SG235Q5, GAUSS1 and SG820Q15. They have been obtained for a signal injected 
with an intrinsic SNR of 7 (see Eq. (\ref{eq:SNR})). 
This SNR value emphasizes the performance differences; with a higher value, for many pipelines
the efficiency saturates at 100\% over the full FAR range.
TABLE \ref{tab:efficiency_50} gives the values of a signals' intrinsic 
SNR necessary to have a detection efficiency of 50\% at a fixed FAR of
0.01Hz. These values have been obtained by fitting the curve of efficiency versus the 
signal $h_{rss}$ with an asymmetric sigmoid function, as defined in \cite{ref:BLS2}.\\

\begin{table}[t]

\begin{tabular}{cc}
\begin{tabular}{l |r r | r r r | r r}
\hline 
H1               & PF   & KW  & QT  & PC  & EGC & MF   & ALF  \\
\hline		  	              	              
A1B2G1           & 6.7  & 7.5 & 6.5 & 5.1 & 5.1 & 7.3  & 6.7  \\  
A2B4G1           & 7.4  & 7.7 & 6.4 & 5.8 & 6.9 & 7.0  & 6.6  \\  
GAUSS1           & 6.2  & 7.0 & 5.5 & 4.9 & 5.6 & 6.7  & 6.1  \\
GAUSS4           & 7.2  & 8.0 & 5.6 & 4.9 & 5.9 & 7.7  & 6.4  \\  
SG235Q5          & 7.7  & 6.9 & 5.1 & 6.1 & 5.8 & 7.5  & 7.0  \\  
SG235Q15         & 10.5 & 8.6 & 5.1 & -   & 6.0 & 10.6 & 9.9  \\  
SG820Q5          & 5.9  & 7.4 & 5.2 & -   & 6.0 & 8.2  & 6.9  \\  
SG820Q15         & 5.6  & 9.1 & 5.1 & -   & 5.8 & 11.6 & 9.7  \\
\hline
\end{tabular}

&

\begin{tabular}{l |r r | r r r | r r}
\hline
V1               & PF   & KW  & QT  & PC  & EGC & MF   & ALF  \\
\hline		  	              	              
A1B2G1           & 5.9  & 7.1 & 6.6 & 5.6 & 5.3 & 6.1  & 6.0  \\  
A2B4G1           & 6.6  & 7.5 & 6.8 & 5.8 & 6.5 & 6.3  & 6.0  \\  
GAUSS1           & 5.9  & 6.5 & 5.9 & 4.9 & 5.5 & 6.0  & 5.7  \\
GAUSS4           & 7.3  & 8.4 & 7.0 & 5.2 & 5.7 & 6.2  & 5.6  \\  
SG235Q5          & 6.6  & 6.9 & 5.2 & 6.6 & 5.5 & 7.5  & 6.9  \\  
SG235Q15         & 8.9  & 8.8 & 5.1 & -   & 5.7 & 11.7 & 9.8  \\  
SG820Q5          & 5.7  & 7.5 & 5.1 & -   & 5.7 & 8.6  & 7.1  \\  
SG820Q15         & 5.7  & 9.3 & 5.3 & -   & 5.6 & 15.9 & 10.2 \\
\hline
\end{tabular}
\end{tabular}
\caption{\label{tab:efficiency_50} Necessary $\rho$ for 50\% efficiency at FAR = 0.01Hz for H1 (left) and V1 (right) data. 
The symbol - means that the performance was too poor to be estimated.}
\end{table} 

The first comment about these numbers is that filters perform differently according to the 
waveforms. There is no one filter that performs better than all of the others when considering 
such a variety of waveforms. 
One can also note that the largest performance spread corresponds to Sine-Gaussian 
signals, as shown in FIG. \ref{fig:rocs}.
QT and EGC have rather constant and good performance for all tested waveforms, 
and as expected perform very well for all Sine-Gaussian waveforms. 
One can, however, remark that QT performs slightly better than EGC on the test-case Sine-Gaussian signals, 
especially for the H1 data. 
This is partly due to the fact that QT is using an over-complete template bank, and also due to different event extraction. 
Moreover, one should note that EGC explores the low frequency region more effectively than QT (the lowest frequency of EGC is 10 Hz 
while QT starts at 64 Hz). That explains why EGC 
obtains better performance on Gaussian signals for the V1 data.

The relative low efficiency of KW with respect for instance to the QT pipeline is partly due to the fact 
that QT is using an over-complete template bank, and to the fact that the Haar function basis does not 
optimally span the region at high Q. 
Moreover, KW has a totally orthogonal basis. The test-case signals SNR is most likely spread over 
several wavelet functions. The KW clustering algorithm tries to recover as much signal as possible, 
but it cannot be as efficient as an optimal matched filter (as extra noise is introduced).

The other filters have less consistent results.
This is especially true for the case of MF, ALF and PC. 
For instance MF, which averages the data for a given 
duration, has a poor (expected and observed) performance on Sine-Gaussian signals. 
For Q constant, MF performance 
decreases as the frequency increases; at a fixed frequency, MF performance decreases as Q increases 
because the number of cycles increases. An equivalent argument explains the poor performance of ALF 
and PC for Sine-Gaussian waveforms. If the intrinsic strength of the signal is spread over many cycles, 
ALF is strongly disabled whatever the size of the analysis window. Conversely, PC and ALF perform
well for signals whose waveform is peak-like (DFM and Gaussian waveforms). PC is, a priori, an optimal 
filter for Gaussian peaks.
PF performance is rather constant; it obtains intermediate performance on DFM and Gaussian signals.
For Sine-Gaussian signals PF is usually very competitive but, as can be seen in FIG. 
\ref{fig:rocs}, this is not the case for the SG235Q5 signal. This loss of efficiency is due to 
the choice of the frequency windows in which the power is calculated. The frequency region around 
235 Hz is split over two windows.

It is also interesting to note some differences in the results obtained with the H1 and V1 data for
signals having the same SNR, as reported in TABLE \ref{tab:efficiency_50}. The difference never exceeds a maximum of
25\%, and is different from one pipeline to another; QT and PC perform better with H1 data 
while the other pipelines perform slightly better on the V1 data. The difference depends also on the 
waveforms; the largest difference is found for the GAUSS4 signal.
The bad performance of QT on the GAUSS4 signal for V1 compared to H1 data is due to the fact that the amplitude 
of the GAUSS4 signal is lower in V1 than in H1 data. Indeed, this signal has its main frequency content at very 
low frequency (below 100 Hz as shown in FIG. \ref{fig:spectra}). The waveforms are normalized for V1 and H1 
such that their SNR is constant in both detectors. 
Since the Virgo PSD is lower at low frequency than LIGO's PSD, the overall amplitude of the GAUSS4  waveform 
is lower for V1 compared to H1. But, since the QT low frequency cut off (64Hz) is the same for V1 and H1, QT will 
recover a smaller fraction of SNR of this signal in V1 compared to H1.

\subsection{Waveform robustness results}
\label{robustness}

In this study, the parameters of the algorithms have been chosen in order to cover 
a large fraction of the parameter space defined in Section \ref{sec:parameter}. 
However, many of the pipelines actually use parameters which have been tuned 
for real data analysis in the LSC or Virgo burst searches. 
The performance of the pipelines, described above, have been estimated using 
eight waveforms spread over the parameter space. 
In order to test the robustness of the pipelines' performance to a wider set of burst waveforms
spread over the signal parameter space, we have 
computed the efficiencies for all pipelines at a given FAR using 100 waveforms chosen randomly as follows:
\begin{itemize}
\item{50 waveforms produced by core collapse simulations (SN) extracted from \cite{ref:zwerger} \cite{ref:dimmelmeier02}} and \cite{ref:ott04}.
\item{50 band passed white noise (WN) signals: white noise is filtered in frequency and then a Gaussian window is applied in the time domain in order to limit the duration of the signal.}
\end{itemize}
The band passed white noise signals have been considered in this study in order to easily span 
the parameter space, especially in the region of high frequency and large frequency bandwidth.
The parameters of the waveforms are given in Appendix \ref{sec:appendixA}.
\begin{figure}[t]
\begin{center}
\begin{tabular}{cc}
\includegraphics[width=9cm,height=8cm]{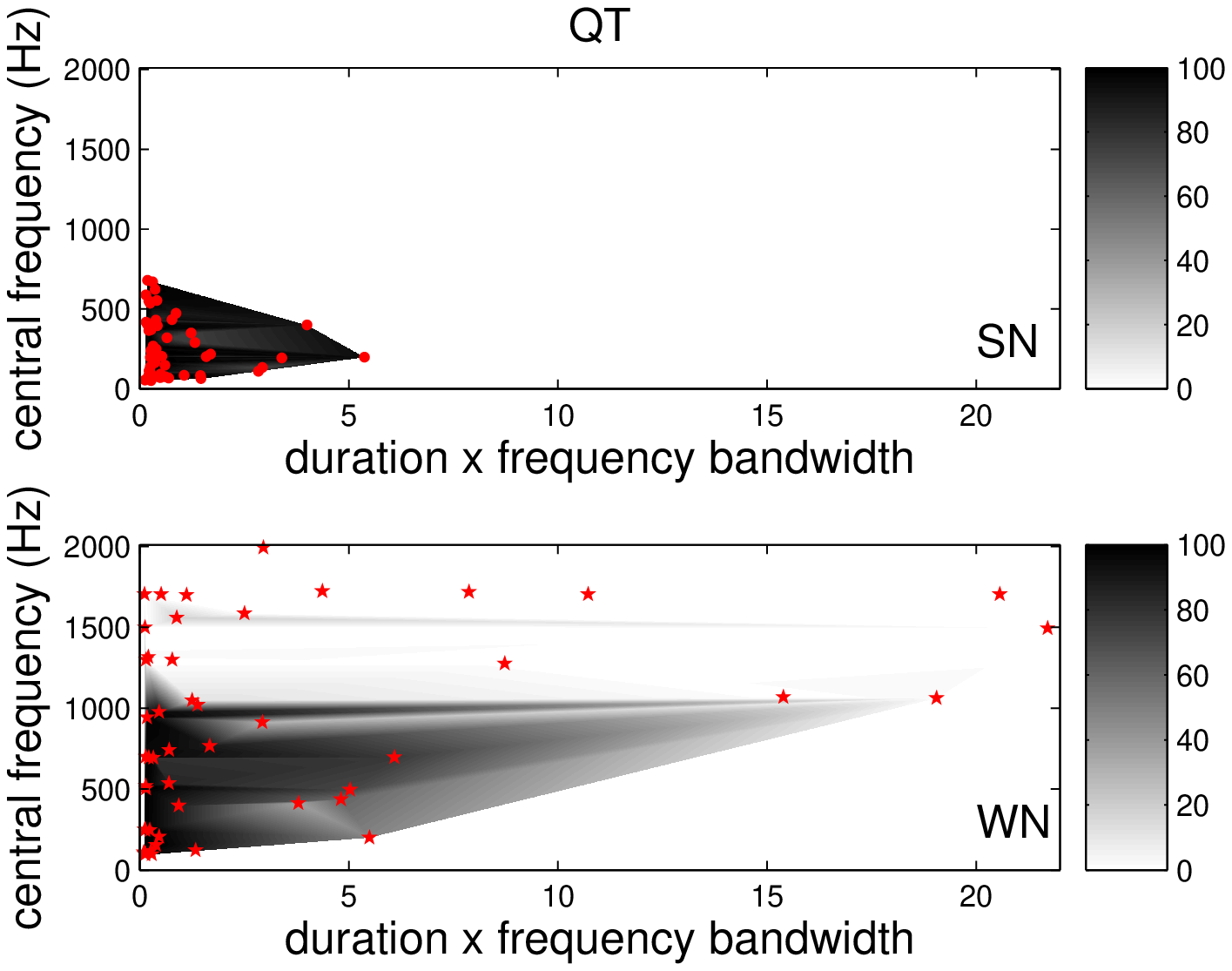} & \includegraphics[width=9cm,height=8cm]{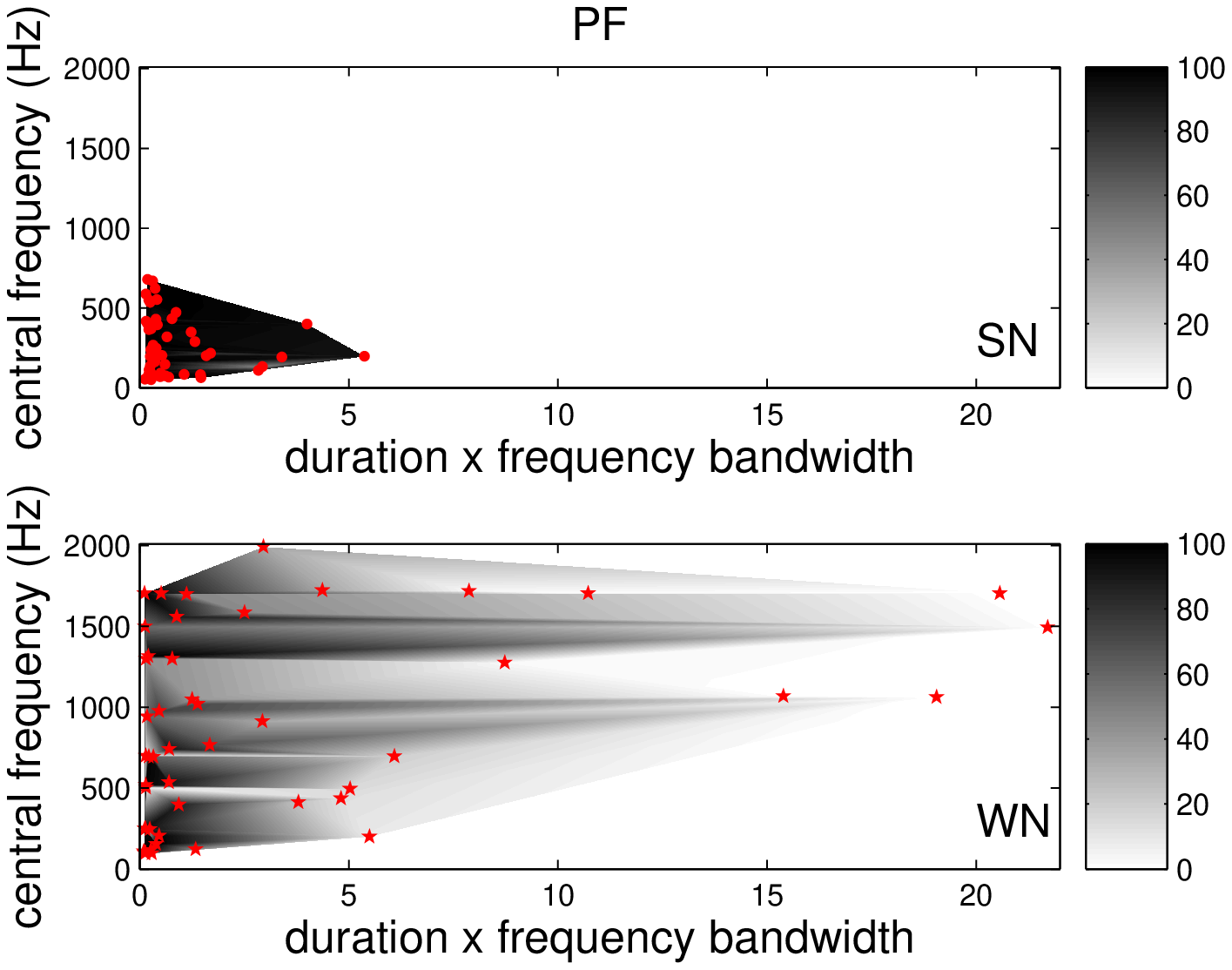}\\
\includegraphics[width=9cm,height=8cm]{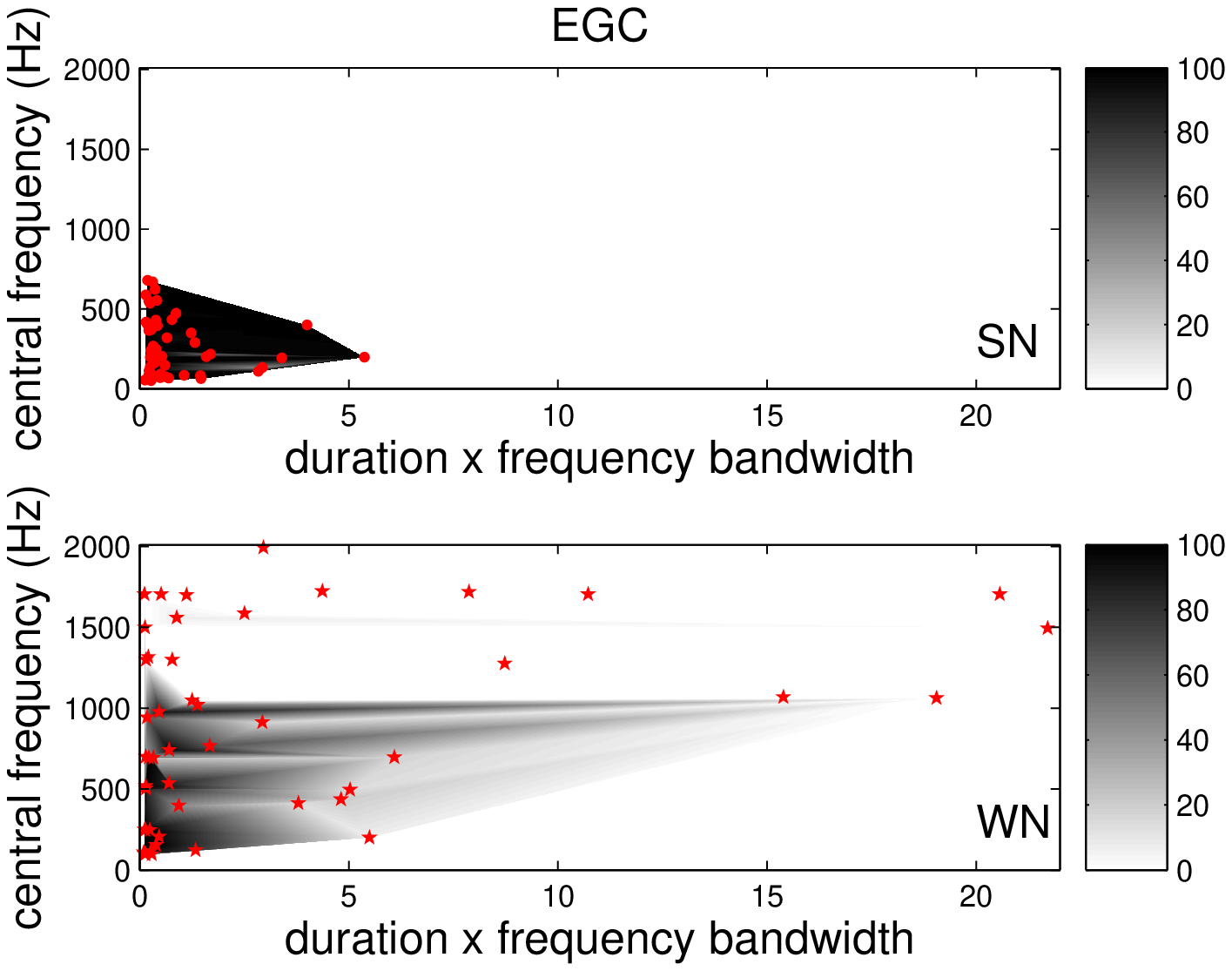} & \includegraphics[width=9cm,height=8cm]{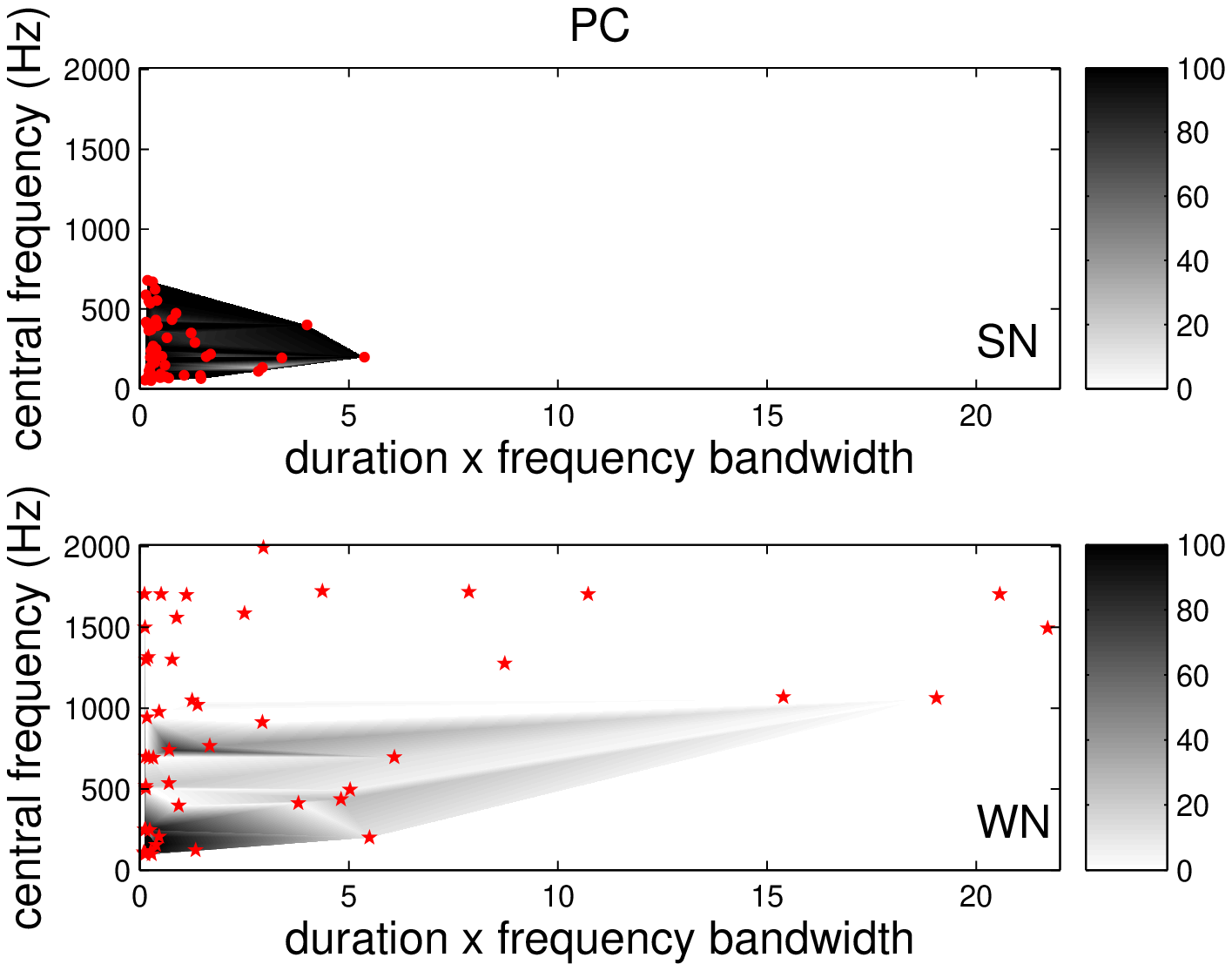}\\
\includegraphics[width=9cm,height=8cm]{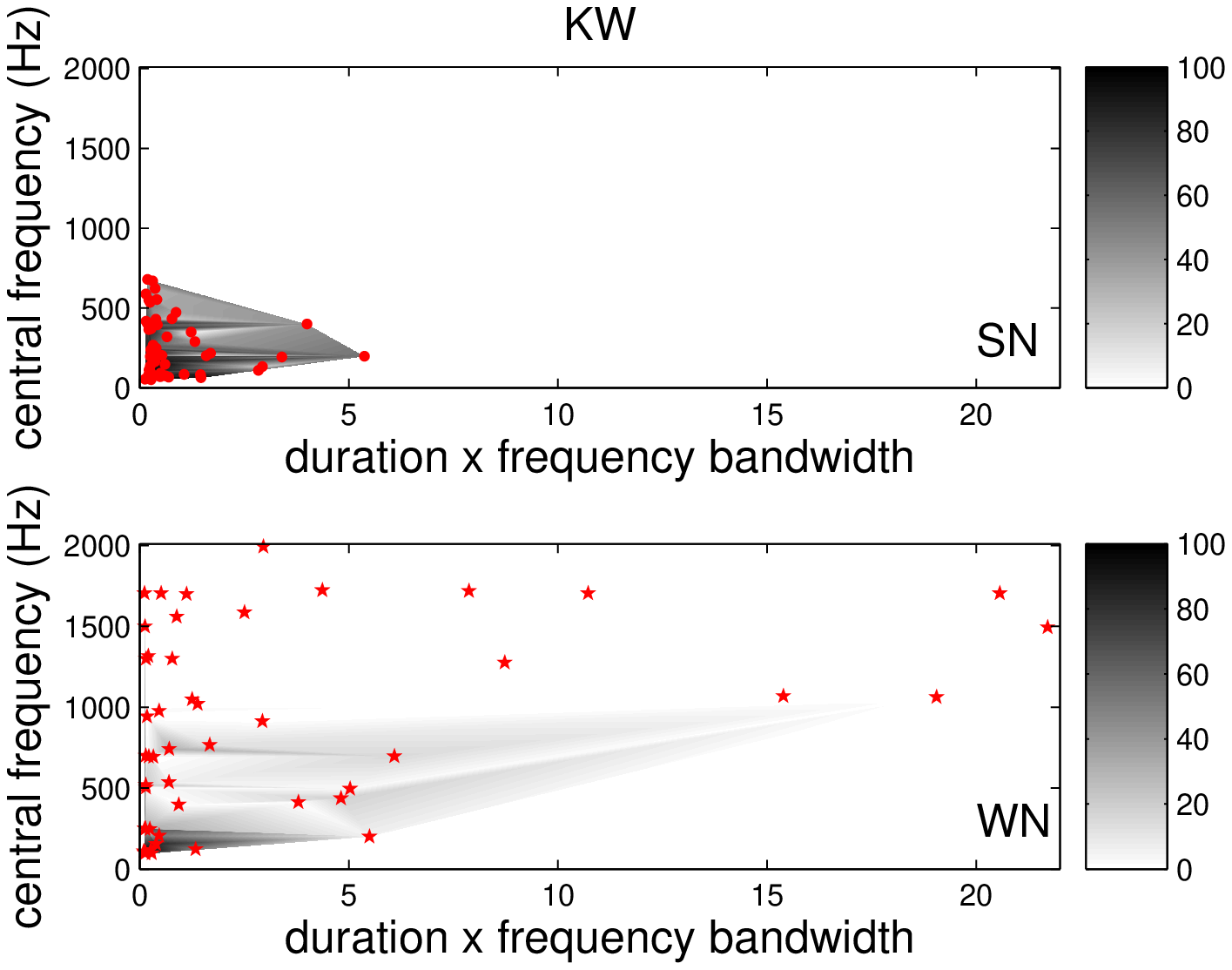} & \includegraphics[width=9cm,height=8cm]{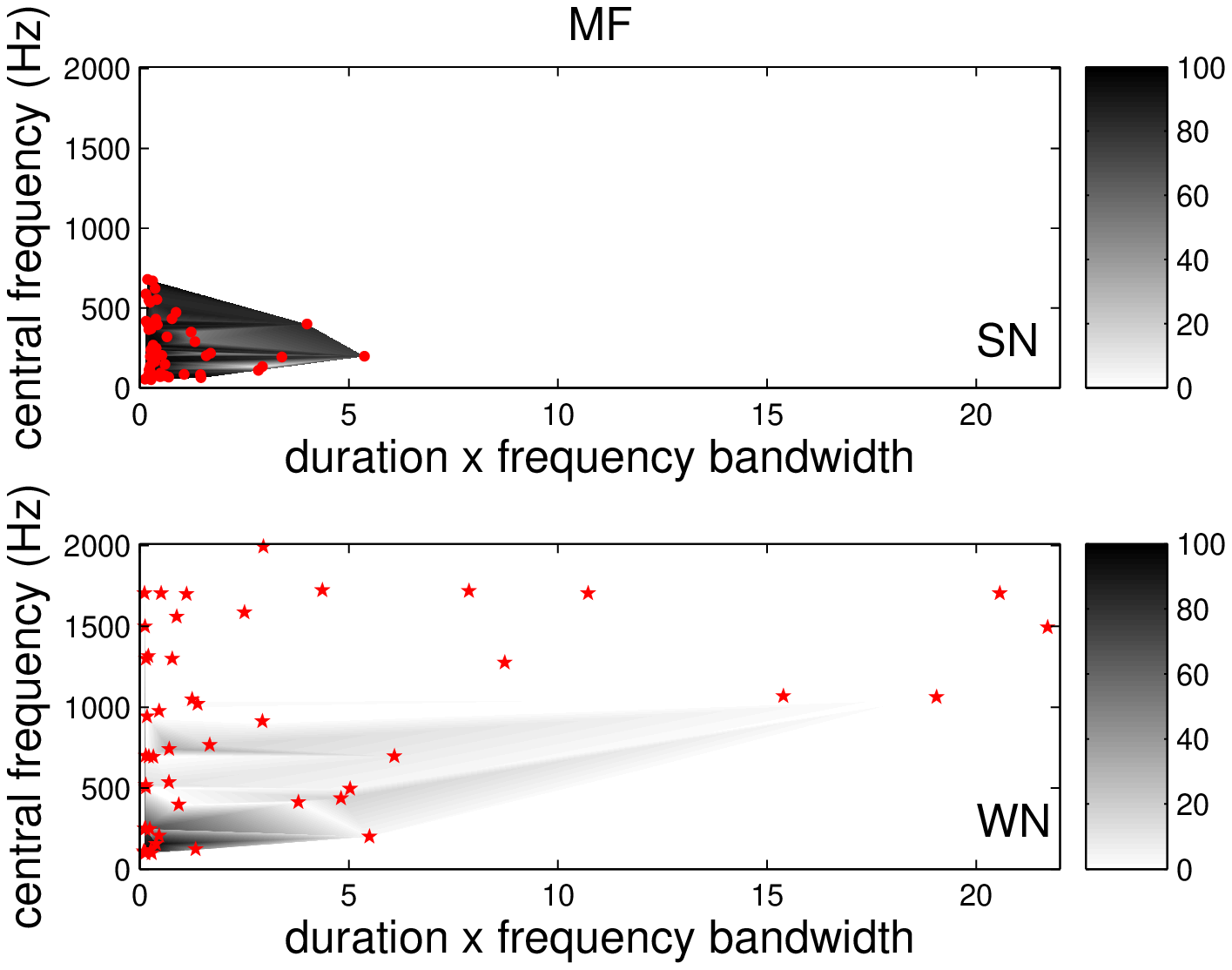}\\
\end{tabular}
\caption{\label{fig:robustness} Pipeline efficiencies obtained for two families of waveforms: SN (top) and WN (bottom).
The efficiency of 6 pipelines (ALF pipeline is not represented here, but it is very similar to MF) is shown in the 2-dimensional plane of 
central frequency versus the time-frequency volume. 
The efficiency has been computed using the Virgo noise data for a FAR 
of 0.001 Hz and a signal SNR of 10.}
\end{center}
\end{figure}
FIG. \ref{fig:robustness} shows the pipelines' efficiency obtained for a FAR of 0.001Hz with the 
V1 data, represented in the two-dimensional plane defined by the central frequency and the 
time-frequency volume (product of the duration with the frequency bandwidth).
An important point concerns the location of the different families of waveforms; although the three 
parameters, based on the moments of the distribution of the signal energy in the time and frequency domains, 
obey uncertainty relation (see Eq. (\ref{eq:heisenberg}) ) that explain the borders of the plots, 
the different signal families do not populate exactly the same region of the parameter space. 
The core collapse signals (SN) are all concentrated in a region for which the central frequency 
is lower than 700 Hz, while the frequency bandwidth can be as large as 600 Hz, 
and the time duration is well spread over the full range.
The band passed WN signals cover the full parameter space, as defined in 
Section \ref{sec:parameter}.
Note that the very low bandwidth WN signals look like Sine-Gaussian signals (a Gaussian window is applied in the 
time domain on all filtered white noise waveforms in order to limit the duration).

One can note that the overall parameter space coverage performances obtained by the different 
pipelines are not identical.
PF is the only pipeline which has been configured to include data in the 1-2 kHz range in its 
analysis. 
Naturally, a pipeline that has been tuned to 
cover the full parameter space has a higher FAR at a given efficiency. This can be translated 
into an efficiency drop of a few percent for a given FAR for all signals.
The performance of the other detection pipelines drops above 1 kHz. It should be mentioned 
that for MF, ALF and PC the tuning is intrinsically limited; the only free parameter of 
these pipelines is the time duration of the signal, thereby limiting the frequency domain 
exploration.  
KW is the filter for which the performance gets worse as the time-frequency volume increases.
Actually, as already mentioned in Section \ref{sec:performance}, because of the choice of the Haar 
function basis the KW performance drops for long duration signals.
One can also note that all pipeline performances are maximum in the region of low or moderate time-frequency volume 
where actual SN signals are located. All the pipelines obtain their best performance on SN signals and fail 
mostly at detecting the WN signals.
Concerning the very low bandwidth signals which look like Sine-Gaussians, QT performs very well as expected, 
while the EGC efficiency is not as good for Sine-Gaussian signals with very low frequency bandwidth. 
EGC is expected to be an optimal filter for Sine-Gaussian, as is QT.
The loss of efficiency is due to the choice of template placement for EGC, which is non-optimal in
the region of the parameter space populated by those particular Sine-Gaussian signals.

\begin{figure}[H]
\begin{center}
\includegraphics[width=12cm]{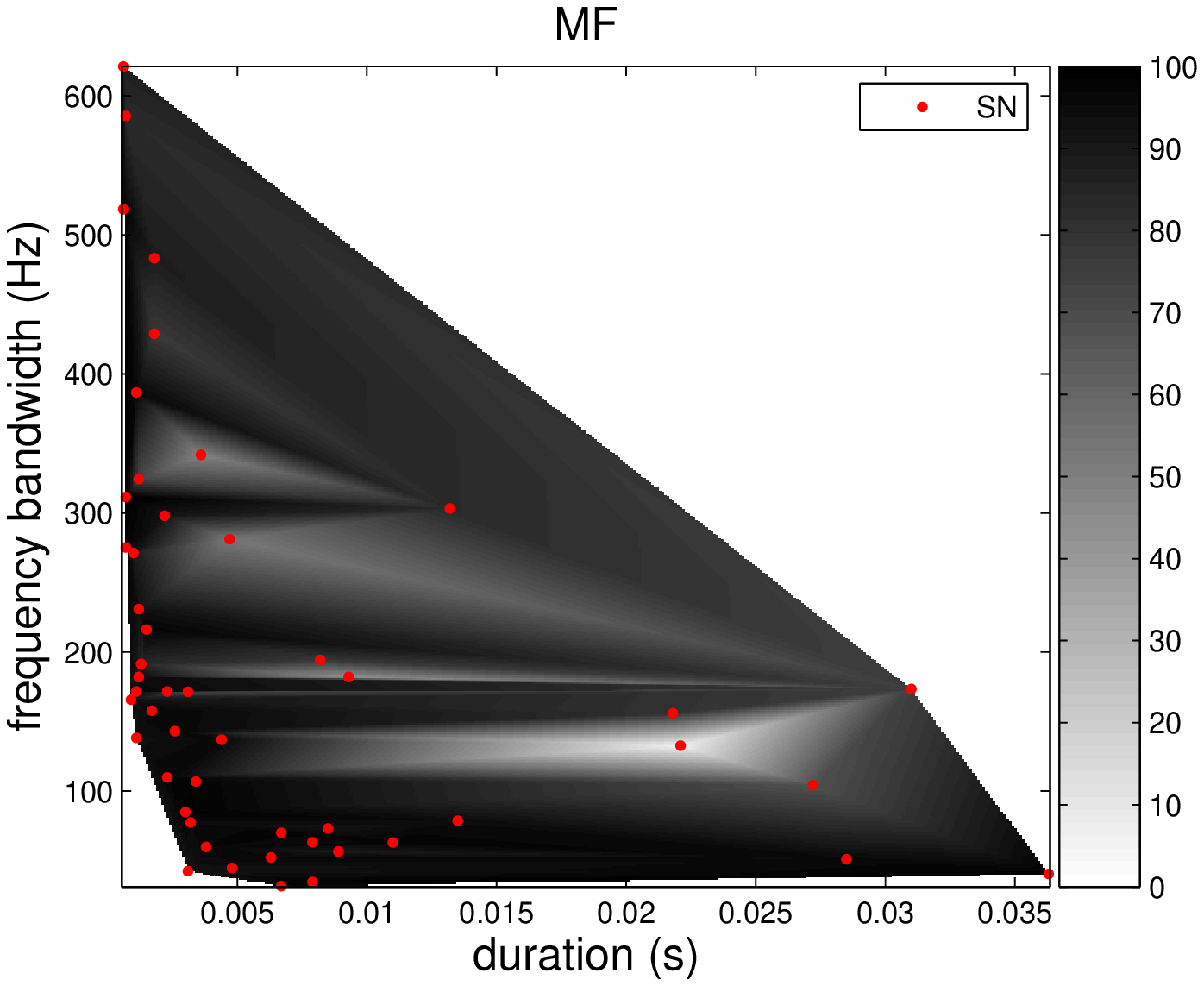} 
\caption{\label{fig:MF-BWvsDURATION} MF pipeline efficiencies obtained for 50 SN waveforms as function of the 
duration and the bandwidth of the signal. The efficiency has been computed for a FAR of 0.001 Hz and the signal
SNR is 10.}
\end{center}
\end{figure}

\begin{figure}[t]
\begin{center}
\includegraphics[width=12cm]{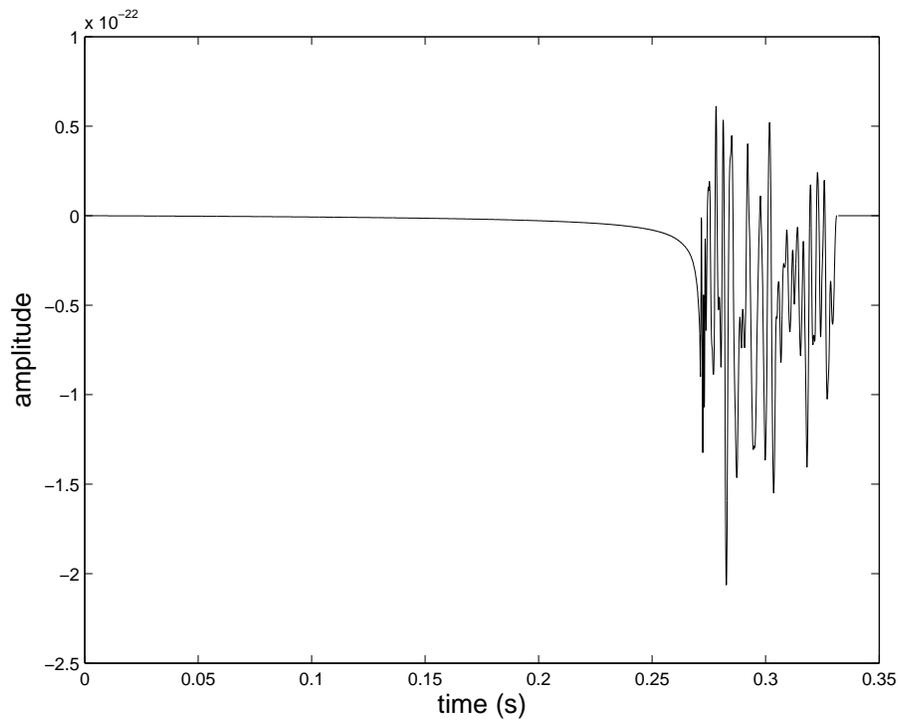} 
\caption{\label{fig:s15nonrot} The waveform of the simulated SN signal (extracted from \cite{ref:ott04} and labelled as s15nonrot in 
Appendix \ref{sec:appendixA}) for 
which the efficiency of all pipelines is low compared to other SN signals close to it in the parameters space defined in this study.}
\end{center}
\end{figure}

Focusing on the SN family, one can see that all pipeline performances drop 
for a particular SN signal, although they all maintain a very high efficiency for the others. 
For instance, the MF efficiency decreases to 6\% for this one SN signal,
while it remains always above 75\% for the 49 other SN signals, and even better for similar signals 
in the parameter space neighborhood, as shown in FIG \ref{fig:MF-BWvsDURATION}.
This efficiency drop is primarily explained by the peculiarity of the shape of 
this core collapse simulated 
signal, which does not show any peak usually associated with core bounce but rather rapid 
oscillations. This particular signal 
corresponds to a rapid collapse for which the first peak usually
present in the SN signals is strongly suppressed (non rotating model). 
The waveform of this SN signal is shown in FIG. \ref{fig:s15nonrot}.
This example tends to prove that the 3 parameters used to describe a gravitational burst event are not complete. 
This fact is even clearer when one considers the WN signal results; 
in the region of the SN signals, the overall performance of all pipelines on WN is lower than for SN signals. 
Moreover for all pipelines, there is a region, inside the SN area (for central frequency between 400Hz and 500Hz), 
where the efficiency drops to a very low value for WN signals,
while there are some SN events with identical parameters
for which the efficiency is close to 100\%. This indicates again that the various pipelines are taking into 
account some signal shape information that might be badly modeled with the 3 parameters used for this study.
But this present study is limited by the number of waveforms tested.

\subsection{Timing accuracy}
\label{sec:timing}
It is crucial to accurately estimate the arrival time of a signal in each gravitational wave detector 
in order to perform an interferometer network coincidence analysis, reconstruct the source 
sky location, and to perform a coincidence analysis with other messengers (neutrinos, gamma ray bursts, etc). 
It is especially important to prove that we are able to
determine the arrival time of a signal in each of the interferometers with an accuracy that is much 
better than the time of flight of a signal between two interferometers (10 ms between the
LIGO interferometers, and 27 ms between Virgo and the LIGO interferometers).
For the bank of burst filters investigated, we have studied the statistical 
properties of the signal arrival time estimators that can be built for each of the filters.
The most natural and convenient definition of the signal arrival time is 
the time for which the amplitude is maximum.
Taking this definition, an estimator for determining the
signal arrival time has been defined based upon the filter statistics. 
Those estimators can be biased and/or can have different intrinsic 
accuracy depending on the waveform of the signal. We report in TABLE \ref{tab:accuracy} the bias 
and the standard deviation of the distribution of the difference between the true arrival time 
and the estimated arrival time obtained for each waveform injected each 60 s
in the V1 3 hour data stream. The signal strength was SNR=10 and the detection efficiency 
was fixed by a common FAR of 0.1Hz for all waveforms. 

As expected the estimator built with the peak correlator filter output (PC) has a very good accuracy 
(less than 0.3 ms) and almost no bias for all signals which concentrate their energy in one peak (Gaussian 
and DFM A1B2G1 signals).
The best accuracy is obtained on the DFM A1B2G1 signal for all estimators. This is due to the
small width of the DFM A1B2G1 peak (width of 0.25 ms, compared with the 2.35 ms 
width of GAUSS1).
Nevertheless, it has not been possible to characterize the arrival time estimator of PC on 
Sine-Gaussian signals due to its very low efficiency.
For the DFM A2B4G1 waveform the bias and standard deviation of all estimators are rather 
large compared to other DFM waveforms due to the presence in the waveform of 3 peaks of decreasing amplitude 
separated by 5 ms. The energy of DFM A2B4G1 is more spread out in the time-frequency plane, 
decreasing the localization ability for all filters.
For Sine-Gaussian waveforms the QT, EGC and PF estimators obtain the best time accuracy; they are the
most efficient filters for detecting these signals.
It is obvious that there is a direct relation between a low efficiency filter and poor time accuracy 
estimator (see for instance PC or MF on Sine-Gaussian waveforms), but an efficient filter does not
necessarily allow one to build a good estimator of the arrival time (see for example ALF on Gaussian peak
waveforms).

\begin{figure}[h]
\begin{center}
\includegraphics[width=12cm]{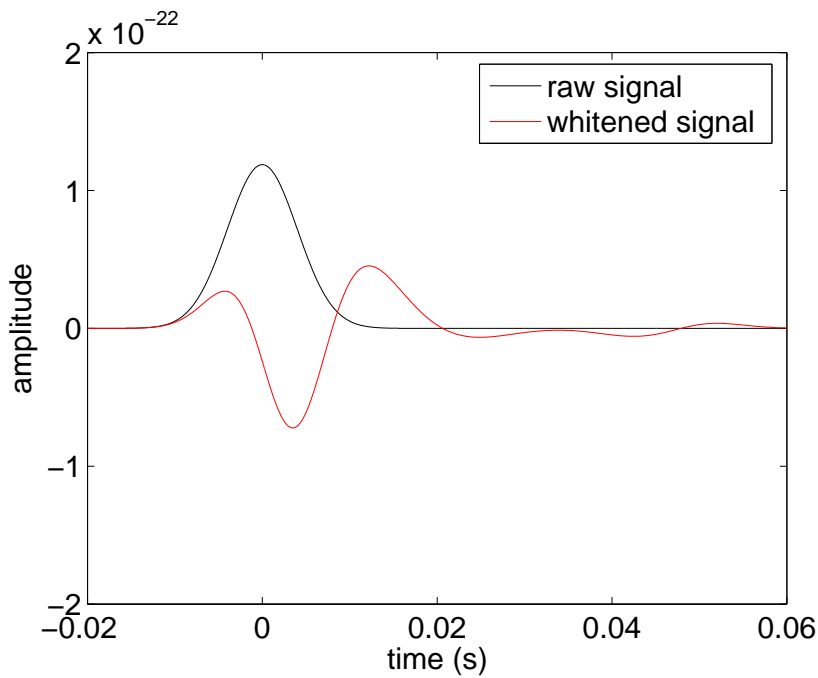}
\caption{\label{fig:gauss4_white} The effect of the V1 whitening pre-processing filter applied on
a Gaussian pulse of 4ms width (GAUSS4).}
\end{center}
\end{figure}

\begin{figure}[h]
\begin{center}
\includegraphics[width=12cm]{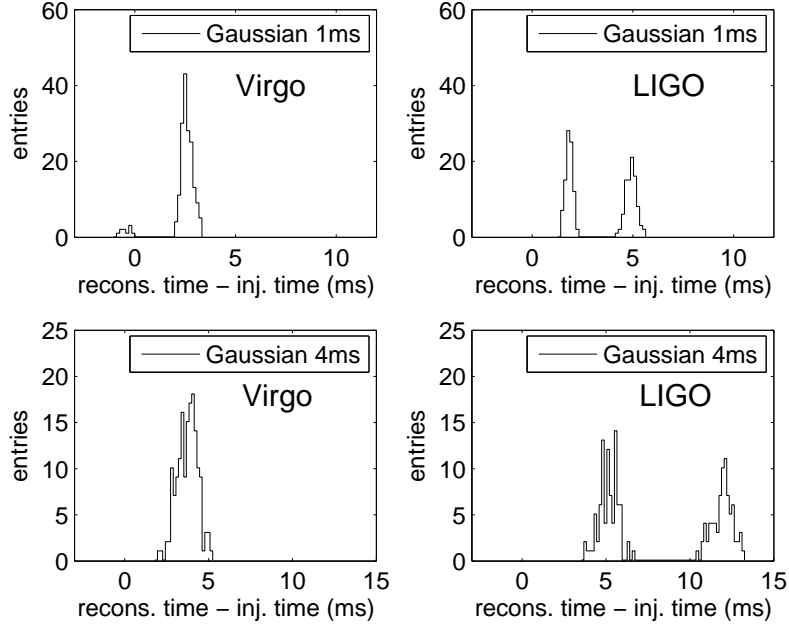}
\caption{\label{fig:reso_ALF_gaussian} Distributions of the difference between the reconstructed 
arrival time of a Gaussian signal (width of 1ms and 4 ms) and the injected value. These 
distributions were created using the MF timing estimator on V1 and H1 data.}
\end{center}
\end{figure}

The large bias (up to 6.4 ms for ALF) of the time domain filters, KW and PF estimators, is explained partly by 
the whitening process applied to the data. Indeed, as shown in FIG. \ref{fig:gauss4_white} for V1, the 
whitening filter strongly distorts a Gaussian waveform. The whitening can transform a Gaussian peak into
a bipolar waveform due to the spectral power shape. As a consequence, the filter triggers on 
one peak or another depending on noise fluctuations. 
The whitening distortion effect is important for the signals that have a large SNR content at 
low frequency, such as the Gaussian signals.
The distortion is even stronger in the case of H1 noise since the power spectrum is higher in 
the low frequency region (the effect of the whitening process is then larger in this region). 
For instance, the MF estimator bias for a 4 ms Gaussian peak increases from 3.4 ms to 8.3 ms and its 
standard deviation from 0.9 ms to 3.4 ms, as shown in FIG. \ref{fig:reso_ALF_gaussian}.
Moreover, the whitening algorithm used by the PF, MF and ALF pipelines was not a non-zero phase filter.
It thus introduces an additional time delay that adds to the bias due to waveform distortion. This delay 
depends on the power spectrum and the waveform.
Finally, it is important to note that for many waveforms the arrival time estimation 
accuracy for a signal of intrinsic SNR of 10 can be as good as 1 ms. The standard deviation
of the arrival time estimation scales roughly inversely with the signal SNR \cite{ref:arnaud03_2}.

\begin{table}
\begin{tabular}{cc}
\begin{tabular}{l |r r | r r r | r r}
\hline
Standard deviation  & PF   & KW   & QT  & PC   & EGC  & MF   & ALF  \\
\hline		   	       	                                
A1B2G1              & 0.05 & 0.5  & 0.2 & 0.04 & 0.03 & 0.05 & 0.3  \\
A2B4G1              & 0.7  & 2.6  & 1.4 & 0.2  & 0.4  & 0.5  & 2.2  \\
GAUSS1              & 1.2  & 1.4  & 0.8 & 0.1  & 0.2  & 0.8  & 1.5  \\
GAUSS4              & 2.7  & 4.3  & 2.3 & 0.3  & 0.7  & 0.6  & 3.2  \\
SG235Q5             & 1.4  & 0.9  & 0.9 & 1.2  & 0.5  & 1.3  & 1.1  \\
SG235Q15            & 4.8  & 3.3  & 2.5 & -    & 1.6  & 4.4  & 4.1  \\
SG820Q5             & 0.2  & 0.3  & 0.2 & -    & 0.2  & 0.3  & 0.3  \\
SG820Q15            & 0.6  & 1.1  & 0.7 & -    & 0.5  & 1.3  & 1.1  \\
\hline
\end{tabular}
&
\begin{tabular}{l |r r | r r r | r r}
\hline
Bias     & PF   & KW   & QT    & PC      & EGC   & MF   & ALF \\
\hline	                       			              
         & 0.2  & 0.3  & -0.1  & -0.05   & -0.06 & 0.03 & 0.1 \\
         & 1.9  & 4.9  & -1.7  & -1.3    & -1.4  & 2.1  & 0.7 \\
         & 1.7  & 3.4  & -0.05 & -0.01   &  0.01 & 2.3  & 1.7 \\
         & 4.1  & 10.7 & -0.4  & -0.03   & -0.03 & 3.4  & 6.4 \\
         & 0.6  & 1.3  & -0.07 &  0.02   &  0.02 & 0.7  & 0.7 \\
         & 1.1  & 0.8  &  0.06 &  -      &  0.1  & 0.2  & 0.6 \\
         & 0.2  & 0.2  & -0.01 &  -      &  0.01 & 0.1  & 0.1 \\
         & 0.2  & 0.3  & -0.04 &  -      &  0.03 & 0.3  & 0.2 \\
 
\hline
\end{tabular}

\end{tabular}
\caption{\label{tab:accuracy} Standard deviation and bias (in ms) of the signal arrival time (in ms) obtained by 
estimators built from the burst filters. The values have been obtained for the different test-case 
waveforms selected at a false alarm rate of 0.1Hz for V1 data. The signal SNR is 10. 
When the efficiency is low, it has not been possible to estimate all the quantities for PC filter estimator.}
\end{table} 

\section{Coincidence analysis}
\label{sec:section4}
\subsection{Description}
\label{sec:coincidence}
As shown in Section \ref{sec:section3}, the FAR of the burst signal filters
will remain quite high if we do not want to miss the rare astrophysical event. For example, a DFM A1B2G1 
signal of $\rho$=6 can be observed with a detection efficiency between 30\% and 75\%, but 
with a 0.01Hz FAR leading to 864 false alarm events per day. 
Furthermore, the FAR for burst searches on real data tend to be much higher than those on Gaussian
stationary data.
One way to significantly reduce the FAR while keeping the detection efficiency 
above, say 50\%, is to request that the arrival times of signals in the 
detectors be compatible with the travel time of the wave between interferometer locations. 
This assumes, of course, that detector noise fluctuations are not correlated. 
It also requires the timing error to be less than the travel time of the wave which is the case for 
all pipelines (except maybe KW for Hanford-Livingston coincidence, see TABLE \ref{tab:accuracy}).
In the case of a source of unknown position, the signal arrival time difference
must be less than the maximum travel time $\Delta_{12}$\, between two detectors.
If the source position is known (for example, a detected optical counterpart) one 
can use the exact travel time of the signal between the detectors. However, this
constraint does not provide a large false alarm reduction. \\

In this study, the source
has been located in the direction of the galactic center, while we are performing
an all-sky blind search using only the timing information (the use of the frequency 
characteristics has been studied separately).
Assuming that the trigger time distribution follows a Poisson distribution, 
the FAR of a coincidence analysis, performed with two interferometers'
output, is given by the product,
$fa_1 . fa_2 . 2\Delta_{12}$, where the $fa_i$~ are the 
FARs in the detector pipelines, and $\Delta_{12}$~ is the maximum 
travel time of a signal from one detector to the other. If we consider a single
experiment FAR of 0.1 Hz, the double experiment coincidence analysis 
FAR is around 0.2 mHz (for $\Delta_{12} = 10$~ms). However, if a detector 
in the network is not well aligned with respect to the others, the signal may be not visible in 
at least one detector, even with a high single interferometer FAR. It is
thus important given the network (Hanford, Livingston and Virgo observatories) to 
study how the different network configurations compare.
Hence we consider coincidence between these three interferometers (three-fold coincidence),
and coincidence between two detectors among the three (two-fold coincidence). 
We also consider the combination of the two-fold coincidences in order to cope with 
the misalignment of the LIGO-Virgo detector network.

\subsection{Two-fold and three-fold analysis}
\label{sec:cointiming}
The seven pipelines have been applied on the three data streams in which six 
different waveforms were injected each minute (the maximum rate to avoid any bias of the
pipeline output estimation due to the presence of too many signals in the data segment). 
The source of given $h_{rss}$\, is located in the direction of the galactic center
and its polarization angle is varied randomly as explained in Section \ref{sec:coininj}. 
We produced event lists corresponding to a 
rate of false alarm events of 0.1 Hz. The efficiencies obtained by the pipelines for the 
six waveforms are reported in TABLE \ref{tab:effi-Ib-ind}.

\begin{table}
\begin{tabular}{l|c|c|c||c|c|c||c|c|c||c|c|c||c|c|c||c|c|c}
\hline
    & \multicolumn{3}{|c||}{A1B2G1} & \multicolumn{3}{|c||}{A2B4G1} & \multicolumn{3}{|c||}{GAUSS1} & \multicolumn{3}{|c||}{GAUSS4} & \multicolumn{3}{|c||}{SG235Q15} & \multicolumn{3}{|c}{SG820Q15} \\ 
\hline
    & H1  & L1  & V1   &  H1  & L1   & V1   & H1  & L1  & V1   & H1  & L1  & V1   & H1  & L1  & V1  & H1  & L1  & V1  \\
\hline
PF      & 42 & 42 & 49 & 49 & 48 & 47 & 66 & 62 & 50 & 28 & 29 & 57 & 35 & 38 & 31 & 42 & 42 & 63 \\
KW      & 37 & 38 & 44 & 50 & 48 & 42 & 61 & 58 & 47 & 23 & 25 & 52 & 50 & 50 & 33 & 19 & 21 & 48 \\
QT      & 46 & 45 & 48 & 59 & 56 & 46 & 72 & 68 & 50 & 40 & 43 & 52 & 71 & 67 & 53 & 48 & 47 & 65 \\
PC      & 55 & 54 & 54 & 65 & 62 & 55 & 77 & 72 & 58 & 53 & 54 & 72 & -    &  -   &  -   &  -   &  -   &  -   \\
EGC     & 57 & 55 & 56 & 54 & 53 & 48 & 72 & 68 & 53 & 40 & 42 & 66 & 67 & 64 & 50 & 41 & 41 & 65 \\
MF      & 40 & 40 & 51 & 53 & 52 & 51 & 64 & 61 & 50 & 27 & 29 & 63 & 38 & 41 & 23 & 10 & 11 & 33 \\
ALF     & 43 & 43 & 52 & 57 & 55 & 52 & 68 & 63 & 53 & 36 & 39 & 68 & 42 & 43 & 25 & 16 & 19 & 44 \\
\hline
average & 46 & 45 & 51 & 55 & 53 & 49 & 69 & 65 & 52 & 35 & 37 & 61 & 43 & 43 & 31 & 25 & 26 & 45 \\
minimum & 37 & 38 & 44 & 49 & 48 & 42 & 61 & 58 & 47 & 23 & 25 & 52 &  0 &  0 &  0 &  0 &  0 &  0 \\
maximum & 57 & 55 & 56 & 65 & 62 & 55 & 77 & 72 & 58 & 53 & 54 & 72 & 71 & 67 & 53 & 48 & 47 & 65 \\
\hline
\end{tabular}
\caption{\label{tab:effi-Ib-ind} Detection efficiencies (in percent) obtained by the seven pipelines for a source 
located in the galactic center for a FAR of 0.1 Hz. The efficiencies have been estimated for each
detector data stream: Hanford (H1), Livingston (L1) and Virgo (V1).
In this study the source emits signals regularly over 24 hours (one emission each minute with random 
polarization angle as explained in Section \ref{sec:coininj}); 
we tested 6 different waveforms. 
The pipelines' efficiencies thus take into account the modulation of the signal strength due to the antenna 
pattern of each detector at a given time. The best, worst and average efficiencies are also given.}
\end{table}

\subsubsection{Coincidence analysis}
Starting from the generated lists of events (FAR is 0.1 Hz), each pipeline looked 
for two-fold and three-fold coincidences. The coincidence time windows applied for a blind search take 
into account the maximum light travel time between pairs of interferometers and the maximum timing 
accuracy (bias and standard deviation) obtained for the test-case waveforms given in TABLE \ref{tab:accuracy}. 
Using the results presented in TABLE \ref{tab:accuracy} we conservatively assumed an error of 10 ms for all pipelines 
and for all signals.
The event compatibility is tested using the estimated arrival time $t_{V}$, $t_{L}$ and $t_{H}$ of an 
event in each of the three detectors.
For the two-fold coincidence analysis, the time difference in one of the 3 considered detector pairs 
must obey:
\begin{eqnarray}
|t_{H} -t_{L}| < \Delta_{HL}  \;\; \textrm{for a H1-L1 two-fold coincidence analysis} \label{eq:HL}\\
|t_{H} -t_{V}| < \Delta_{VHL} \;\; \textrm{for a H1-V1 two-fold coincidence analysis} \label{eq:HV}\\
|t_{L} -t_{V}| < \Delta_{VHL} \;\; \textrm{for a V1-L1 two-fold coincidence analysis} \label{eq:LV}
\end{eqnarray}
where $\Delta_{HL} = 20$ ms and $\Delta_{VHL} = 38$ ms are the time windows of coincidence for the H1-L1 pair, and 
for the H1-V1 or L1-V1 detector pairs respectively.
For the three-fold coincidence analysis, the conditions (\ref{eq:HL}), (\ref{eq:HV}) and 
(\ref{eq:LV}) must be fulfilled simultaneously.\\

\subsubsection{FAR estimation}
It has been verified that the FAR does not depend on the pipeline 
but only depends on the coincidence time window that is applied for the 
given network configuration following the formula given in Section \ref{sec:coincidence}. 
This supports the hypothesis that the event triggers for all pipelines are distributed according 
to a Poisson distribution.
The FAR has been estimated precisely for each pipeline using the trigger list obtained
from noise;
in order to cope with the expected low number of false alarm events per day,
especially in the case of the three-fold coincidence (most of the pipelines end up with 0 or 
1 false alarm event), we have increased artificially by 10000 times the duration of the simulation. 
We do this by time shifting, by a step of 1-second, the event arrival time obtained from noise data streams. 
In practice, for a two-fold coincidence analysis, we consider the two trigger lists and add to the arrival time
of all the events of one list an offset of 1 second. 
We perform coincidence and determine the number
of coincident events and then we repeat the operation but adding an offset of 2 seconds, etc.
Note that the events of a N-second time-shifted stream that are in the last N seconds are moved 
to the beginning of the time-shifted stream such that the duration of the two streams in coincidence is always identical 
to 24 hours. 
A 1-second time shift is long enough (the typical event duration is much smaller than 1 second) to 
be sure of having different triggers list each time we add an additional 1-second offset. We checked that similar results 
are obtained with a longer time offset.
For the three-fold analysis, a similar action is done on two of the 3 trigger lists; a 1-second offset is added
to one of the trigger lists, and a 2-second offset is added to the other. That guarantees that 
at each step we perform coincidence analysis with trigger lists as if we had simulated new data streams.\\

For a two-fold coincidence search the FAR 
is $(7.5 \pm .1)\, 10^{-4}$\, Hz for a pair involving Virgo ($7.6 \times 10^{-4}$\, Hz expected), and 
$(4.03 \pm .03)\, 10^{-4}$\, Hz for the Livingston-Hanford pair ($4.1 \times 10^{-4}$\, Hz expected). 
For the three-fold coincidence the FAR is $(3.07 \pm .03)\, 10^{-6}$\, Hz ($3.04 \times 10^{-6}$\, 
Hz expected).

\subsubsection{Two-fold, three-fold and combined two-fold analysis}
In TABLE \ref{tab:effi-Ib-double} and \ref{tab:effi-Ib-triple} we give the detection efficiencies 
obtained for the two-fold coincidence and the three-fold coincidence analysis respectively. 
It is interesting to note that for a given waveform the three-fold coincidence efficiencies of the 
pipelines differ significantly. The maximum (30 \%) and average (20\%) values obtained for GAUSS1
are low due to the non-optimum alignment 
of the three detectors. On the other hand, we end up with a FAR which is easily manageable, with 
less than 1 false alarm per day. 
This number is not achievable by any individual interferometer search.
However, requiring coincidence among three interferometers that 
are not optimally aligned might not be the best strategy, 
compared to a two-fold coincidence search using Hanford and Livingston, which are well-aligned.
Note that, as expected, the efficiency of the Livingston-Hanford pair is always higher than any pair involving
Virgo. Among the two other pairs, Virgo-Hanford is the worst due to the least optimal
antenna response overlap.
The results of the two-fold coincidence, given in TABLE \ref{tab:effi-Ib-double}, indicate 
that on average the efficiency can be up to 70\% higher than in the three-fold coincidence case, but 
the FAR remains high (a few $10^{-4}$\, Hz).
We thus estimate the efficiency of the two-fold coincidence analysis at the FAR of the 
three-fold coincidence analysis to be $10^{-6}$~ Hz. 
The results are summarized in TABLE \ref{tab:effi-Ib-double-micro}.
The two-fold efficiency is higher than the three-fold efficiency (comparison done at the same FAR)
and, of course, the Hanford-Livingston pair is always favored compared to any pair including the Virgo
detector.
One can also require the detection of events in one of the three pairs of interferometers. The main
advantage of this strategy is to reduce the FAR because of the coincidence constraint, 
while recovering events which are located in a sky zone for which only two detectors are sensitive.
The results are summarized in TABLE \ref{tab:effi-Ib-double-combined} for all pipelines. 
The combined two-fold coincidence analysis has been done using event lists obtained for a FAR of $10^{-6}$ in each pair of detectors, such that the combined two-fold analysis FAR is $3 \times 10^{-6}$ Hz; it is the sum of the FAR of the two-fold coincidence from which the events that are actually
common to the 3 pairs of interferometers must be subtracted two times. This number of events is given by the
three-fold FAR which is at least 3 orders of magnitude lower than the two-fold FAR, and then can 
therefore be neglected; 
this can be directly compared to the three-fold coincidence FAR (which was used to make TABLE \ref{tab:effi-Ib-triple}) 
of $3 \times 10^{-6}$ Hz. 

\begin{table}
\begin{tabular}{l|c|c|c||c|c|c||c|c|c||c|c|c||c|c|c||c|c|c}
\hline
    & \multicolumn{3}{|c||}{A1B2G1} & \multicolumn{3}{|c||}{A2B4G1} & \multicolumn{3}{|c||}{GAUSS1} & \multicolumn{3}{|c||}{GAUSS4} & \multicolumn{3}{|c||}{SG235Q15} & \multicolumn{3}{|c}{SG820Q15} \\ 
\hline
    & HL   & HV   & LV    &  HL  & HV   & LV    & HL   & HV   & LV   & HL   & HV   & LV    &   HL & HV   & LV    &  HL  & HV   & LV   \\
\hline
PF      & 29 & 15 & 17  & 34 & 17 & 18  & 48 & 28 & 27 & 18 & 12 & 14  & 24 &  5 &  9  & 28 & 22 & 22    \\
KW      & 25 & 10 & 13  & 34 & 15 & 17  & 43 & 22 & 24 & 13 &  9 & 10  & 35 &  9 & 14  & 11 &  4 &  6    \\
QT      & 31 & 16 & 18  & 40 & 22 & 22  & 53 & 32 & 31 & 28 & 16 & 19  & 53 & 33 & 32  & 33 & 27 & 26    \\
PC      & 38 & 25 & 25  & 50 & 32 & 30  & 59 & 42 & 38 & 38 & 35 & 36  &  -   &  -   &  -    &  -   &  -   &  -      \\
EGC     & 40 & 27 & 27  & 38 & 21 & 21  & 54 & 35 & 33 & 28 & 21 & 24  & 50 & 29 & 29  & 28 & 22 & 22    \\
MF      & 26 & 14 & 17  & 37 & 22 & 22  & 47 & 27 & 27 & 17 & 14 & 16  & 28 &  4 &  9  &  4 &  1 &  3    \\
ALF     & 29 & 17 & 18  & 40 & 25 & 24  & 50 & 32 & 30 & 23 & 20 & 22  & 31 &  6 & 11  &  9 &  4 &  7    \\
\hline
average & 31 & 18 & 19  & 39 & 22 & 22  & 51 & 31 & 30 & 24 & 18 & 20  & 32 & 12 & 15  & 16 & 11 & 12 \\
minimum & 25 & 10 & 13  & 34 & 15 & 17  & 43 & 22 & 24 & 13 &  9 & 10  &  0 &  0 &  0  & 0  &  0 &  0 \\
maximum & 40 & 27 & 27  & 50 & 32 & 30  & 59 & 42 & 38 & 38 & 35 & 36  & 53 & 33 & 32  & 33 & 27 & 26 \\
\hline
\end{tabular}



\caption{\label{tab:effi-Ib-double} Detection efficiencies (in percent) obtained from the seven pipelines for a source 
located in the galactic center for a double coincidence search between a pair of interferometers. The efficiency 
has been estimated for each combination: Hanford-Livingston (HL), Hanford-Virgo (HV) and Livingston-Virgo (LV).
The two-fold coincidence analysis has been done using events list obtained for a FAR of 0.1 Hz 
in each detector. The FAR of each two-fold coincidence involving Virgo is $7.5 \times 10^{-4}$ Hz and $4 \times 10^{-4}$ Hz for the 
Hanford-Livingston two-fold coincidence analysis (for all pipelines).}
\end{table}

\begin{table}
\begin{tabular}{l|c|c|c|c|c|c}
\hline
    & A1B2G1  & A2B4G1 & GAUSS1  & GAUSS4  & SG235Q15 & SG820Q15 \\ 
\hline
PF  &  9   & 10    & 17    &  7    &   3   & 13   \\
KW  &  6   &  9    & 13    &  4    &   6   &  2   \\
QT  &  9   & 12    & 21    &  9    &  22   & 17   \\
PC  & 14   & 19    & 30    & 23    &  -      &  -     \\
EGC & 16   & 12    & 23    & 13    &  19   & 13   \\
MF  &  8   & 13    & 17    &  8    &   3   &  -   \\
ALF &  9   & 14    & 20    & 12    &   3   &  2   \\
\hline
average & 10 & 13 &  20   &  11   &  8  &  7 \\
minimum &  6 &  9 &  13   &   4   &  0  &  0    \\
maximum & 16 & 19 &  30   &  23   & 22  & 17    \\
\hline
\end{tabular}



\caption{\label{tab:effi-Ib-triple} Detection efficiencies (in percent) obtained from the seven pipelines for a source 
located in the galactic center for a three-fold coincidence search between the Hanford, Livingston and Virgo 
interferometers.
The coincidence analysis has been done using events list obtained for a FAR of 0.1 Hz 
in each detector. The three-fold coincidence FAR is $3 \times 10^{-6}$ Hz for all pipelines.}
\end{table}


\begin{table}
\begin{tabular}{l|c|c|c||c|c|c||c|c|c||c|c|c||c|c|c||c|c|c}
\hline
    & \multicolumn{3}{|c||}{A1B2G1} & \multicolumn{3}{|c||}{A2B4G1} & \multicolumn{3}{|c||}{GAUSS1} & \multicolumn{3}{|c||}{GAUSS4} & \multicolumn{3}{|c||}{SG235Q15} & \multicolumn{3}{|c}{SG820Q15} \\ 
\hline
    & HL   & HV   & LV    &  HL  & HV   & LV    & HL   & HV   & LV   & HL   & HV   & LV    &   HL & HV   & LV    & HL  & HV   & LV    \\
\hline
PF      & 23 &  9 & 11 & 29 & 10 & 13 & 43 & 18 & 20  & 13  &  5  &  7  & 15  &  1 &  4 & 22 & 12 & 15  \\
KW      & 19 &  5 &  9 & 27 &  7 & 11 & 38 & 12 & 16  &  9  &  3  &  5  & 29  &  4 &  9 &  7 &  2 &  4  \\
QT      & 25 &  8 & 12 & 35 & 12 & 16 & 48 & 22 & 22  & 22  &  9  & 12  & 48  & 25 & 25 & 26 & 16 & 18  \\
PC      & 33 & 15 & 18 & 39 & 20 & 20 & 53 & 32 & 30  & 30  & 23  & 24  & -     & -    & -    & -    & -    &  -     \\
EGC     & 35 & 18 & 20 & 31 & 11 & 14 & 47 & 24 & 24  & 20  & 13  & 15  & 43  & 19 & 21 & 21 & 10 & 14  \\
MF      & 20 &  7 & 10 & 30 & 12 & 15 & 40 & 15 & 18  & 10  &  6  &  8  & 18  &  -   &  3 &  -   &  -   &  -  \\
ALF     & 23 &  8 & 12 & 34 & 14 & 16 & 43 & 20 & 22  & 16  & 11  & 14  & 21  &  1 &  4 &  4 &  -   &  2  \\
\hline
average & 25 & 10 & 13 & 32 & 12 & 15 & 45 & 20 & 22  & 17  & 10  & 12  & 25  &  7 &  9 & 11 &  6 & 8 \\
minimum & 19 &  5 &  9 & 27 &  7 & 11 & 38 & 12 & 16  &  9  &  3  &  5  &  0  &  0 &  0 &  0 &  0 & 0 \\
maximum & 35 & 18 & 20 & 39 & 20 & 20 & 53 & 32 & 30  & 30  & 23  & 24  & 48  & 25 & 25 & 26 & 16 & 18 \\
\hline
\end{tabular}

\caption{\label{tab:effi-Ib-double-micro}Detection efficiencies (in percent) obtained by the seven pipelines for a source 
located in the galactic center for a double coincidence search between pairs of interferometers. The two-fold 
coincidence analysis has been done using event lists compiled such that the two-fold coincidence false alarm 
rate is $10^{-6}$~ Hz. The efficiency has been estimated for each combination: Hanford-Livingston (HL), 
Hanford-Virgo (HV), and Livingston-Virgo (LV).}
\end{table}

\begin{table}
\begin{tabular}{l|c|c|c|c|c|c}
\hline
HL $\cup$ HV $\cup$ LV    & A1B2G1  & A2B4G1 & GAUSS1  & GAUSS4  & SG235Q15 & SG820Q15 \\ 
\hline
PF      & 33  & 39   & 60    & 19    & 19   & 34 \\
KW      & 26  & 36   & 52    & 14    & 38   & 11 \\
QT      & 35  & 48   & 66    & 32    & 67   & 40 \\
PC      & 56  & 67   & 76    & 55    &  -     &  -   \\
EGC     & 52  & 43   & 66    & 32    & 60   & 32 \\
MF      & 28  & 43   & 55    & 17    & 20   &  -   \\
ALF     & 33  & 47   & 60    & 28    & 25   &  6 \\
\hline
average & 38  & 46   & 62    & 28    & 33   & 18 \\
minimum & 26  & 36   & 52    & 14    &  0   &  0  \\
maximum & 56  & 67   & 76    & 55    & 67   & 40  \\
\hline
\end{tabular}

\caption{\label{tab:effi-Ib-double-combined}Detection efficiencies (in percent) obtained by the seven pipelines for a source 
located in the galactic center requiring that at least one pair of interferometers among the three detects 
the signal. This combined two-fold coincidence analysis has been done using event lists obtained for a 
FAR of $10^{-6}$~ Hz in each pair of detectors, such that the combined two-fold analysis FAR is $3 \times 10^{-6}$ Hz 
identical to the three-fold analysis.}
\end{table}

\subsection{Coincidence searches using the frequency information}
Some of the burst pipelines (QT, PF, KW and EGC) are able to estimate the characteristics of the events 
in the frequency domain. Specifically, these are the frequency bandwidth (defined by $f_{min}$\, and $f_{max}$) 
and the peak frequency ($f_{peak}$) which corresponds to the template or wavelet coefficient that has the highest SNR.
The peak frequency is thus not necessarily in the middle of the frequency bandwidth.
Those coincidence searches whose results are given in Section \ref{sec:cointiming} considered only the 
timing information. The QT pipeline used in the LSC is actually performing a combined coincidence 
search requesting timing and frequency coincidence. It has been shown \cite{ref:QT} that the QT 
coincidence search pipeline can reach a better efficiency (of a few percent) for the same FAR when using 
the frequency information. 
It is worth mentioning that the time and frequency coincidence performed in the LSC QT pipeline is done before the event
extraction; this results in pruning overlapping templates. In this study the time and frequency coincidence is done
using event lists.\\

In this study, we tried to estimate the benefit of adding the frequency information to a coincidence search.
For this, we used the trigger lists generated for each interferometer at 0.1 Hz, and applied the same 
timing coincidence plus a constraint on the overlap of the bandwidth of the triggers: 

\begin{eqnarray}
                     f_{max}^i > f_{min}^j    \;\;  \textrm{if} \;\; f_{min}^i < f_{min}^j \nonumber \\
\textrm{or} \; \;    f_{max}^j > f_{min}^i    \;\;  \textrm{if} \;\; f_{min}^i > f_{min}^j
\end{eqnarray}
where $i$~ and $j$~ are the index of two of the three detectors.
We estimated the efficiencies of the three pipelines on 
the test-case signals for a two-fold and the three-fold coincidence searches and then compared them to the results given 
in TABLE \ref{tab:effi-Ib-double} and \ref{tab:effi-Ib-triple} (time only coincidence). 
The efficiencies obtained were very similar (with a loss lower than 1\%) to those obtained for the time 
only coincidence search. 
The FAR is reduced, but in a rather moderate way depending on the pipelines: a factor of 0.67 
for EGC and KW, a factor of 0.4 for PF, and a factor of 0.2 for QT for the three-fold coincidence search. 
The FAR reduction factors are given in TABLE \ref{tab:FAR-Ib-TimeFrequency} for the different network 
configurations.

\begin{table}
\begin{tabular}{c|ccc|c}
\hline
network                          & VH            & VL            & HL            & VHL           \\
\hline
FAR (time only) QT PF KW and EGC & $7.6 \times  10^{-4}$ & $7.6 \times 10^{-4}$ & $4.0 \times 10^{-4}$ & $3.0 \times 10^{-6}$ \\

\hline
FAR (time and frequency) QT      & $3.2 \times 10^{-4}$ & $3.2 \times 10^{-4}$ & $1.7 \times 10^{-4}$ & $4.9 \times 10^{-7}$ \\
Reduction factor QT              & .4              & .4              & .4              & .2            \\

\hline
FAR (time and frequency) PF      & $5.0 \times 10^{-4}$ & $5.0 \times 10^{-4}$ & $2.6 \times 10^{-4}$ & $1.2 \times 10^{-6}$ \\
Reduction factor PF              & .7              & .7              & .7              & .4            \\

\hline
FAR (time and frequency) KW      & $6.2 \times 10^{-4}$ & $6.2 \times 10^{-4}$ & $3.4 \times 10^{-4}$ & $2.0 \times 10^{-6}$ \\
Reduction factor KW              & .8              & .8              & .9              & .7            \\

\hline
FAR (time and frequency) EGC     & $6.2 \times 10^{-4}$ & $6.2 \times 10^{-4}$ & $3.7 \times 10^{-4}$ & $2.2 \times 10^{-6}$ \\
Reduction factor EGC             & .8              & .8              & .9              & .7            \\
\hline

\end{tabular}
\caption{\label{tab:FAR-Ib-TimeFrequency} FAR (in Hz) obtained by the four burst pipelines that estimate 
the frequency characteristics of the events in this study (QT, PF, KW and EGC). 
The results are given for a time-only 
coincidence search and for the time-and-frequency coincidence search. In the latter case, one requests 
that the frequency bandwidth of events overlaps.}
\end{table}

In addition, we have also tried to use the peak frequency information instead of the bandwidth in order
to determine if events are coincident in the frequency domain. 
The event frequency compatibility test is then:

\begin{equation}
|f_{peak}^i - f_{peak}^i| < \Delta f
\end{equation}
where $\Delta f$~ has been varied such as the efficiency remains constant compared to a time-only coincidence 
analysis. The FAR reduction was found to be slightly 
higher compared to the bandwidth overlap coincidence constraint. This is especially the case for PF, for which 
the FAR of a three-fold coincidence is reduced by a factor 0.08 compared to the time-only coincidence, 
keeping for all test-case waveforms the same efficiency.
But for the other pipelines (QT, KW and EGC), the signal efficiency drops by at least 10\% absolute
for some of the waveforms (mainly the large bandwidth signals such as A1B2G1 or GAUSS1).
Indeed, for large bandwidth signals, these pipelines fail to estimate precisely the peak frequency,
which hence seems not to be a robust enough parameter to be used for a time-and-frequency coincidence 
pipeline (to maintain the efficiency to the time-only coincidence value, it was mandatory to increase 
$\Delta f$~ such that no FAR reduction was observed); an exception to this is the PF pipeline.

\subsection{Comparisons and detection strategy discussion}
In the previous sections we have estimated the efficiency of seven pipelines for different search strategies: 
three-fold, two-fold, and combined two-fold coincidence. We have first shown that the performance of some of 
the pipelines depends strongly on the waveform. The extreme case is reached by the correlator PC
which performs very well for the DFM and Gaussian waveforms, but fails at detecting Sine-Gaussian waveforms.
On the other hand, the EGC and QT pipelines' performance remains very high for all kinds of waveforms. 
The PF and KW pipelines have similar performance whatever the waveform, but perform
on average a bit less efficiently.

The two-fold coincidence results show that one can decrease the FAR from 0.1 Hz (8640 events per day)
to a few $10^{-4}$ Hz (30 events per day) while maintaining a detection efficiency at an acceptable level. 
The best performances are, of course, obtained from the Hanford-Livingston pair; these efficiencies are, on average for 
all pipelines, .15 lower than the individual interferometer pipeline efficiencies working at a 0.1 Hz FAR.
The efficiency o fthe HV or LV combination is approximately 2/3 of the HL efficiency, due to the different alignment of LIGO and Virgo.
To reduce significantly the FAR, one has to request a three-fold coincidence; we then end up with 2 false
alarm events per week. However, due to the non-optimum alignment of the three interferometers the detection 
efficiency drops to 10-20\% on average for the test-case waveforms. To overcome the problem of the 
non-optimum alignment of the three interferometers, while keeping a very low number of false alarm events, the best 
strategy seems to be such as to request the detection of an event in at least one of the three pairs of 
detectors (combined two-fold coincidence).
We have seen that, for the same FAR (2 events per week), the efficiency is 3 times higher, on average, 
compared to a three-fold coincidence search.
The best performing pipeline efficiency is then between 40\% and 76\%, depending on the waveform for 2 
false alarm events per week.

Moreover, comparing the two-fold coincident search obtained with the Hanford-Livingston pair to the combined 
two-fold coincidence, we have shown that, on average for all 
waveforms, the detection efficiency is increased by 50\% by adding Virgo to the Hanford-Livingston network 
(we have checked that the two-fold coincidence Hanford-Livingston efficiencies for a FAR of $3 \times 10^{-6}$~ Hz 
are very similar to those given in TABLE \ref{tab:effi-Ib-double-micro} for a FAR of $10^{-6}$~ Hz.
The difference is at maximum as high as 0.02).
If we consider the results of the two-fold combined search in TABLE \ref{tab:effi-Ib-double-combined}, 
QT and EGC are the two best performing pipelines over the six test-case waveforms. The others have the main 
drawback of performance depending on the waveforms. 

In addition, the figure of merit of the time-only coincident search can be improved if one requires
that coincident events have similar frequency characteristics. We have seen that the reduction of the false 
alarm events actually depends, from pipeline to another, on the frequency estimation of the event bandwidth.
One can expect to reduce the false alarm event number by up to a factor 0.08 without losing any efficiency
in the most favorable case (PF pipeline).

\section{Combination of filters}
\label{sec:section5}
In Section \ref{sec:section3}, the detection potential of several pipelines has been compared using
a few test-case waveforms. We have seen that a pipeline's performance is different from one
waveform to another, and there is the indication that the pipelines are complementary and can 
cover the full range of tested waveforms.
Conversely, for a given signal, it could be interesting to determine if all pipelines detect 
the same events. We would expect a gain of the detection efficiency by combining different
filters' output. Of course, merging (OR analysis) the trigger lists of different pipelines increases the FAR. 
If the pipelines all detect the same events, it would then be more interesting to request that 
the filters' combination keep only the events seen by all the considered filters (AND analysis). 
But in that case, the detection probability will always be dominated by the filter that is least sensitive.
We could however expect that the pipelines are not sensitive to the same noise events and hence see a reduction of the FAR.
In this study, we have considered the 120 possible combinations given 7 pipelines, from 2 filters up to the 7 filter combination.
To fairly compare the results we need to derive the ROC curves and compare them to the single pipeline curves.

We started by using the seven pipelines' trigger lists provided for the eight test-case waveforms injected 
into the V1 and H1 data sets (optimal orientation) with a SNR of 5 and 10, and for 3 FAR 
(0.1Hz, 0.01Hz and 0.005 Hz). 
We chose identical FAR for all filters for the sake of simplicity, while they might not correspond to the optimal choice.
Next, from the single method trigger lists we produced a new set of triggers; 
a single gravitational wave burst candidate was created via the merger of triggers that were overlapping or 
temporally closer than 0.1 seconds. 
This condition was introduced in order to account for the maximum timing error of the filters without merging triggers
due to noise with real signal ones (we start from triggers lists computed at a maximum FAR of 0.1 Hz).
The non-merged triggers are simply added to the new list for the OR analysis.
These prescriptions allowed us to compute both the FAR for the combination of triggers as well as the efficiency. 
Then, for each type of combination (2 filters, 3 filters, etc) we keep the one having the best ROCs. In such a way we can, 
at the end, compare the 6 combined ROCs with the 7 pipelines' ROC.

\begin{figure}[h]
\begin{center}
\includegraphics[width=12cm]{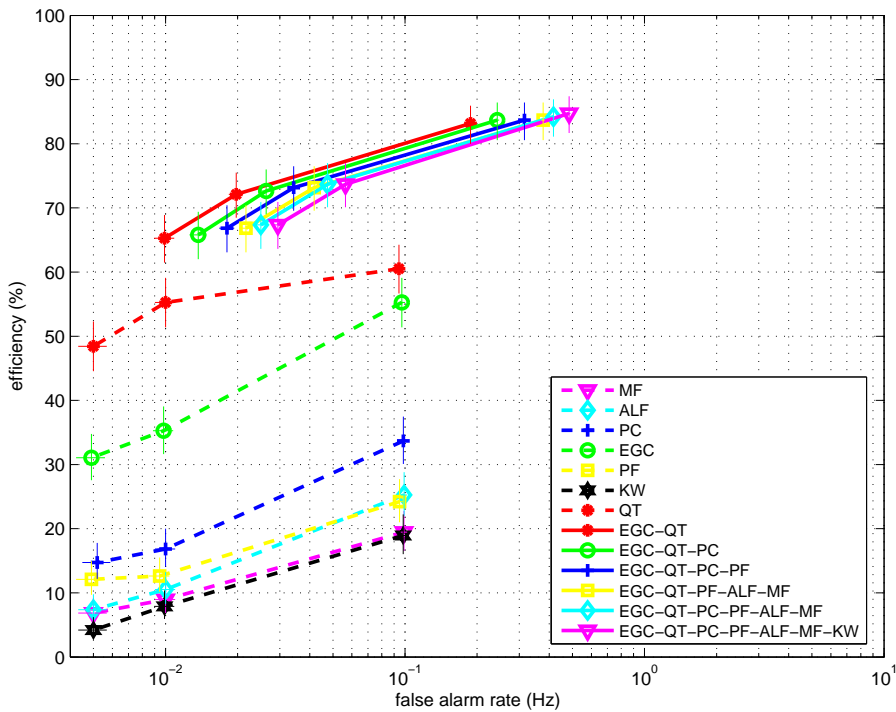}
\caption{\label{fig:q5f235_OR} Three operating points for single methods and the best 
performing combination of methods in the V1 data for an injected Sine-Gaussian SG235Q5 with a SNR of 5. The OR analysis
obtained its best performance on this Sine-Gaussian test-case waveform. 
The vertical error bars have been computed with a binomial fit 
and the horizontal ones by Poisson statistics. 
It can be noted that adding new methods to the EGC-QT combination only translates the ROC horizontally. 
This means that in this case adding methods increases the number of false alarms but not the number of observed injections.}
\end{center}
\end{figure}

The main conclusion of this analysis is that no combination shows a ROC consistently better than the best single method ROC, 
for all waveforms. As expected the AND analysis helps at reducing the FAR, but the efficiency decrease is too 
large to provide any real gain. The result is robust with respect to the injection signal SNR and 
the choice of interferometer. Occasionally for a specific waveform, like the Sine-Gaussian SG235Q5 
(as shown in FIG. \ref{fig:q5f235_OR}), the OR combination of the 2 best pipelines mildly improves the ROC; 
the relative efficiency gain is between 10\% and 30\% depending on the FAR. 
But, this improvement depends on the waveform; for the  A1B2G1 and GAUSS4 signals the performance 
is as good as the best pipeline, or even worse.
In FIG. \ref{fig:q5f235_OR}, we can also note that actually QT performs best on Sine-Gaussian SG235Q5, and combining
QT with EGC increases the detection efficiency from 55\% (QT only) to 65\% (QT or EGC) at a FAR of $10^{-2}$~ Hz.
Since these results do not encourage the combination of methods we decided not to pursue it for 
the network study where only combinations of interferometers' configurations have been thoroughly studied.

\section{Source sky location reconstruction}
\label{sec:section6}
\subsection{Introduction}
In addition to the increase of the detection probability of a burst gravitational
wave, the other significant advantage of the LIGO and Virgo detector network is to provide
the possibility to estimate the position of the source on the sky. 
It is well known \cite{ref:tourrenc} that a minimum of three detectors is 
mandatory in order to determine source sky position, even if an ambiguity remains between two positions symmetric with 
respect to the plane defined by the 3 detectors. So, the LIGO-Virgo network 
is the first step toward gravitational wave astronomy that includes source localization. 
The source coordinates on the sky are right ascension ($\alpha$) and 
declination ($\delta$).
A method \cite{ref:VSL} has been developed for estimating the source position 
using only the arrival time $t_i$ of the gravitational wave in each detector, and 
its associated arrival time error $\sigma_i$. 
This method has the advantage of being fast and independent of the pipeline. 
Three-fold coincidence search events found by the seven pipelines in the 24 hours
of data with a FAR of $3 \times 10^{-6}$ Hz have been considered in this sky location study. 
A comparison of the source reconstruction accuracy of each pipeline is 
presented below.

\subsection{Method} 
Each burst pipeline provides a list of three-fold coincidence events 
defined by the arrival time and the event SNR reconstructed in the three 
detectors. We then have to estimate the arrival time error for each event. 
In this study, for the sake of simplicity, we have taken the following relation 
that is exact for a Gaussian signal of half-width 1 ms detected by PC \cite{ref:arnaud03_2}:

\begin{equation}
  \label{eq:sigma_t}
  \sigma_i = \frac{1.4}{SNR_i} ~~\mbox{ms}
\end{equation}

It should be noted that, in general, the burst pipelines can 
determine, event by event, the error of the arrival time, for instance, by taking into account the
results given in Section \ref{sec:timing}. We have checked that the results weakly depend on the error
parametrization given in (\ref{eq:sigma_t}) as the result is primarily constrained by the arrival time values.
To estimate $\alpha$ and $\delta$ a least-square minimization has been used. Our $\chi^2$ 
function is defined by:

\begin{equation}
  \label{eq:chi2}
  \chi^2 = \sum_{i=1}^{n} \frac{\left(t_i - (t_0 + \Delta_i^{Earth}(\alpha,\delta) ) \right)^2}{\sigma_i^2}
\end{equation}
where $t_0$ is the arrival time of the gravitational wave at the center of the Earth, and
$\Delta_i^{Earth}(\alpha,\delta)$ is the delay between the center of the Earth and
the i$^{th}$ detector at a given sidereal time (and only depends on $\alpha$ and $\delta$).
First of all, this definition of the $\chi^2$ function allows the fitting
of uncorrelated data while the usual reconstruction using two time
differences between detector implies variables with correlated errors deserving a
more difficult treatment.
By introducing $t_0$, there is no reference detector D$_1$ used to compute timing differences. 
The second advantage is that the network is not limited to three detectors, and the addition 
of other detectors is straightforward.
Nevertheless, this method requires that the event be seen by all the detectors above a given
threshold, whereas coherent methods do not need this first selection step.

\subsection{Results} 
\begin{figure}[h]
\begin{center}
\includegraphics[width=12cm]{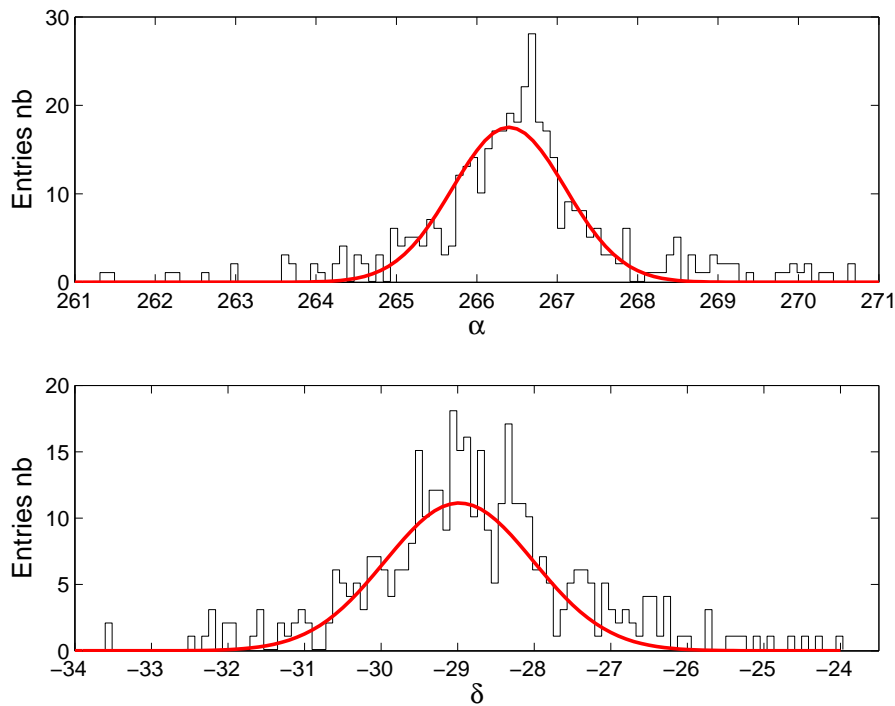}
\caption{\label{fig:reco_gauss1_pc} Distribution of source coordinates $\alpha$ and $\delta$ (in degrees) obtained 
for a GAUSS1 waveform detected with the PC pipeline in the LIGO-Virgo network. The source has been 
placed in the direction of the galactic center and emits signals regularly over 24 hours. The curve is the result of 
a Gaussian fit of the data.}
\end{center}
\end{figure}

FIG. \ref{fig:reco_gauss1_pc} shows an example of the reconstructed source coordinate 
distributions using the 24 hours of simulated data. 
It has been obtained using 456 GAUSS1 events detected by the PC pipeline. 
These distributions include the effect of modulation due to the earth rotation over 24 hours,
which then induces some variation of the event arrival time errors.
The errors on $\alpha$ and $\delta$ are about 1 degree (respectively 0.7$^\circ$ and 1$^\circ$).
TABLE \ref{tab:reco_allfilters} gives a summary of angular reconstruction obtained 
for each pipeline on all waveforms studied.
As expected, the estimation accuracy is mainly dependent on the arrival time measurement accuracy.
In this case the correlator pipelines (PC and EGC) and QT produce the best results.
The achieved precision depends also on the considered waveform. The best accuracy is obtained for 
waveforms which have rather simple shape, such as clearly defined peak (like DFM A1B2G1 or GAUSS1). 
On the contrary, when the waveforms become more complex (DFM A2B4G1, Sine-Gaussian) all the 
pipelines that make very few assumptions on the exact waveform (MF, ALF, PF, KW) fail at 
precisely locating the source. 
QT and EGC perform, as expected, the best for Sine-Gaussian signals.
One should also note that pipelines having a large timing bias have a poor source reconstruction 
estimation, due to the fact that the bias is mainly different in each detector. We have seen 
that bias is predominantly due to the use of non-zero phase whitening filtering. Since LIGO and Virgo
have different power spectrum shapes the bias is different. 
That tends to prove that the source position determination should be somehow decoupled from the detection 
problem. 
Once good candidates are detected, techniques including more information (for instance rough waveform estimation 
as needed by Markov Chain Monte Carlo algorithms \cite{ref:inspiral}) should be used to estimate the source position. 

\begin{table}
\begin{tabular}{c|c|c|c|c|c|c|c}
\hline
       &          & DFM A1B2G1             & DFM A2B4G1           & GAUSS1         & GAUSS4         & SG235Q15           & SG820Q15 \\
\hline
       & $\alpha$ & 265.0 $\pm$ 4.0  & 260.3 $\pm$ 1.9   & 263.0 $\pm$ 7.0  & 250.0 $\pm$ 14.0  & 258.0 $\pm$ 29.0 & 266.0 $\pm$ 5.0  \\
PF     & $\delta$ & -28.0 $\pm$ 7.0  & -27.0 $\pm$ 8.0   & -29.0 $\pm$ 13.0 & -13.0 $\pm$ 21.0  & -15.0 $\pm$ 33.0 & -28.0 $\pm$ 10.0 \\
       & \#       & 126              & 138               & 255              & 104               & 37               & 184               \\

\hline
       & $\alpha$ & 265.0 $\pm$ 5.0  & 264.7 $\pm$  9.0   & 265.0 $\pm$ 8.0  & 265.0 $\pm$ 22.0  & 264.0 $\pm$ 29.0 & 266.0 $\pm$ 6.0  \\
KW     & $\delta$ & -26.5 $\pm$ 9.0  & -26.0 $\pm$ 13.0   & -25.0 $\pm$ 12.0 & -29.0 $\pm$ 23.0  & -15.0 $\pm$ 32.0 & -27.0 $\pm$ 10.0 \\
       & \#       & 93               & 159                & 227              & 61                & 102              & 38                \\
\hline
       & $\alpha$ & 266.0 $\pm$1.5   & 264.0 $\pm$ 6.0     &  266.0 $\pm$ 4.0   & 268.0 $\pm$ 13.0  &  266.0$\pm$ 11.0   &  267.0 $\pm$ 3.0 \\
QT     & $\delta$ & -28.8 $\pm$4.6   & -28.0 $\pm$ 12.0    &  -28.5 $\pm$ 8.0   & -25.0 $\pm$ 17.0  &  -25.0$\pm$ 16.0   &  -29.0 $\pm$ 6.0 \\
       & \#       & 134              & 179                 & 300                & 131               &  318               & 238 \\
\hline
       & $\alpha$ & 266.5  $\pm$ 0.5 & 266.7 $\pm$ 1.3   & 266.4 $\pm$ 0.7  & 266.6 $\pm$ 2.0   &                  &   \\
PC     & $\delta$ & -29.0  $\pm$ 0.6 & -29.1 $\pm$ 2.3   & -29.0 $\pm$ 1.0  & -29.1 $\pm$ 3.6   &        -         & - \\
       & \#       & 240              & 273               & 456              & 334               &                  &   \\
\hline
       & $\alpha$ & 266.3 $\pm$ 0.3  & 265.6 $\pm$ 3.0   & 266.2 $\pm$ 1.0  & 266.7 $\pm$ 3.0   & 265.0 $\pm$ 9.0  & 267.1 $\pm$ 3.0 \\
EGC    & $\delta$ & -28.9 $\pm$ 0.4  & -29.2 $\pm$ 7.0   & -29.0 $\pm$ 1.7  & -28.1 $\pm$ 6.0   & -28.7 $\pm$ 13.0 & -29.0 $\pm$ 4.5 \\
       & \#       & 223              & 172               & 338              & 193           ?    & 269              & 191             \\

\hline
       & $\alpha$ & 266.5  $\pm$ 0.5 & 261.3 $\pm$ 1.3   & 263.7 $\pm$  8.0 & 258.5 $\pm$ 18.0  & 263.0 $\pm$ 30.0 &   \\
MF     & $\delta$ & -29.0  $\pm$ 0.6 & -27.3 $\pm$ 4.0   & -30.2 $\pm$ 14.0 & -10.0 $\pm$ 22.0  & -24.0 $\pm$ 29.0 & - \\
       & \#       & 112              & 182               & 243              & 115               & 41               &   \\ 
\hline
       & $\alpha$ &  265.5 $\pm$ 1.8 &  258.0 $\pm$ 14.0 & 260.1 $\pm$ 1.7  & 258.0 $\pm$ 3.0   & 260.0 $\pm$ 23.0 & 264.0 $\pm$ 10.0 \\
ALF    & $\delta$ &  -27.0 $\pm$ 5.0 &  -25.9 $\pm$ 20.0 & -29.4 $\pm$ 7.0  & -17.0 $\pm$ 10.0  & -27.0 $\pm$ 28.0 & -28.0 $\pm$ 10.0 \\
       & \#       & 132              & 208               & 288              & 175               & 50               &  32 \\

\hline

\end{tabular}
\caption{\label{tab:reco_allfilters}Reconstructed source coordinates obtained by the burst pipelines for the different 
waveforms studied. Standard deviation are also given. 
The galactic center is located at $(266.4^{\circ},-28.98^{\circ})$. The number of three-fold coincident
events used to determine the sky position is also given.}
\end{table}

\section{Summary}
\label{sec:section7}
In this study we have compared different pipelines developed both in the LSC and Virgo for the search for
unmodeled gravitational wave bursts. For both Virgo and LIGO data sets, we have evaluated the pipeline performance
and validated their ability to detect a signal. Among all the seven pipelines studied we have especially 
shown that some are more efficient than others. 
We have demonstrated the benefits of a LIGO-Virgo combined search for the detection of a signal; adding Virgo to the 
Hanford-Livingston interferometer network can increase by 50\% the detection efficiency (without increasing the false alarm rate).
We have shown that the combined two-fold coincident search is 3 times more efficient than a three-fold 
coincident search (for the same expected number of false alarm events).
By contrast, combining the filters' trigger lists obtained on the same data stream does not improve
the detection potential of the burst analysis.
The second advantage of the LIGO-Virgo network is its capability to locate a source. We have shown that in the most favorable case
we can expect to reconstruct the source position with a precision of the order of 1 degree for both the 
right ascension and the declination. The position determination performance is not as good when the waveform is complex in structure.
It would be interesting, in the near future, to compare the performance obtained by the combined two-fold 
coincident search to coherent searches \cite{ref:klimenko} using the same network of interferometers, 
as well as their respective performance to reconstruct the source position \cite{ref:coh89}.

The next step of this work is to analyze test sets of real data from LIGO and Virgo. 
This will bring forward other issues to be studied, 
such as the examination and handling of the quality of the data in a combined search.
One of the key issues will be the non-stationarity of the data sets (the increase of the noise during short 
periods, for instance due to seismic activity) that can produce higher 
FAR than what has been optimally considered in this study.
That could have an impact especially on the combined two-fold 
coincidence analysis FAR. 
Another aspect of the GW burst search are the glitches present in real interferometer data. 
If a careful analysis is undertaken in order to understand the origin of the glitches, and thereby allowing one to suppress most of them, then 
dedicated techniques to distinguish real GW burst signal from noise transients could be used \cite{ref:chatterji}.

\acknowledgments
LIGO Laboratory and the LIGO Scientific Collaboration gratefully acknowledge
the support of the United States National Science Foundation for the construction 
and operation of the LIGO Laboratory and for the support of this research. 
N.C. also acknowledges the support of the Fulbright Scholar Program. 

\appendix
\section{Parameters used for the robustness test}
\label{sec:appendixA}

TABLE~\ref{tab:parameter_rob} provides the parameters associated with the waveforms used in this study. 
These are the signals that were used to test the coverage of the GW parameter space by the burst search pipelines in Section \ref{robustness}.

\begin{table}[h]
\begin{tabular}{l|c|c|c||l|c|c|c}
\hline
name & peak freq. (Hz) & bandwidth (Hz) & duration (s) & name & peak freq. (Hz) & bandwidth (Hz) & duration (s) \\
\hline
    DFM (New.) A1B2G1     &     365.3    &       182.1 &     0.0012  &  WN   &   155.3   &       171.6 &     0.0023\\
    DFM (New.) A4B1G1     &     429.5	 &       324.4 &     0.0012  &  WN   &   200.6   &       176.0 &     0.0312\\
    DFM (New.) A4B5G5     &     138.1	 &        84.9 &     0.0030  &  WN   &   101.1   &        21.7 &     0.0059\\
    DFM (Rel.) A3B1G1     &     552.5	 &       585.5 &     0.0007  &  WN   &    97.2   &        10.4 &     0.0125\\
    DFM (Rel.) A3B2G1     &     433.3	 &       428.9 &     0.0018  &  WN   &   100.6   &         8.1 &     0.0175\\
    DFM (Rel.) A3B2G2     &     622.3	 &       621.1 &     0.0006  &  WN   &    97.0   &        10.8 &     0.0272\\
    DFM (Rel.) A3B4G2     &      77.9	 &        52.3 &     0.0063  &  WN   &   110.1   &        43.8 &     0.0023\\
    DFM (Rel.) A4B4G5     &      67.3	 &        63.1 &     0.0110  &  WN   &   121.6   &        78.8 &     0.0170\\
    DFM (Rel.) A1B3G2     &     670.1	 &       518.4 &     0.0006  &  WN   &   118.3   &        99.5 &     0.0027\\
    DFM (Rel.) A4B5G4     &      69.7	 &        63.2 &     0.0079  &  WN   &   206.1   &       193.9 &     0.0024\\
    DFM (Rel.) A4B1G1     &     288.8	 &       281.2 &     0.0047  &  WN   &   250.2   &        43.1 &     0.0028\\
    DFM (Rel.) A4B1G2     &     349.8	 &       341.8 &     0.0036  &  WN   &   248.3   &         7.6 &     0.0186\\
    DFM (Rel.) A3B3G2     &      83.7	 &        78.7 &     0.0135  &  WN   &   247.2   &        55.7 &     0.0044\\
    DFM (Rel.) A4B2G2     &     201.1	 &       194.3 &     0.0082  &  WN   &   398.0   &        39.6 &     0.0236\\
    BO s11A1000B0.3       &     197.0	 &       109.9 &     0.0023  &  WN   &   413.6   &       200.8 &     0.0189\\
    BO s11A50000B0.6      &      59.0	 &        31.6 &     0.0067  &  WN   &   437.0   &       400.9 &     0.0120\\
    BO s11A500B0.4        &      92.2	 &        56.6 &     0.0089  &  WN   &   495.2   &       320.6 &     0.0157\\
    BO s11A1000B0.4       &      81.6	 &        44.7 &     0.0048  &  WN   &   503.6   &        11.0 &     0.0137\\
    BO s15A500B0.4        &      74.8	 &        73.0 &     0.0085  &  WN   &   503.0   &         9.0 &     0.0170\\
    BO s15A1000B0.3       &     239.8	 &       157.7 &     0.0017  &  WN   &   519.3   &        41.6 &     0.0035\\
    BO s15A1000B0.6       &      51.3	 &        34.6 &     0.0079  &  WN   &   535.7   &       118.9 &     0.0059\\
    BO s15nonrot          &     133.4	 &       132.7 &     0.0221  &  WN   &   766.4   &       541.2 &     0.0031\\
    BO s20A50000B0.1      &     417.1	 &       138.2 &     0.0011  &  WN   &   696.4   &       553.8 &     0.0110\\
    BO s20A500B0.2        &     202.2	 &       171.4 &     0.0031  &  WN   &   699.7   &         7.6 &     0.0193\\
    BO s20A50000B0.2      &     401.7	 &       171.5 &     0.0011  &  WN   &   695.2   &        85.4 &     0.0026\\
    BO s20A50000B0.5      &     267.0	 &       216.2 &     0.0015  &  WN   &   693.8   &        16.2 &     0.0204\\
    BO s20nonrot	  &     217.0	 &       181.9 &     0.0093  &  WN   &   742.3   &       443.6 &     0.0016\\
    BO s20A1000B0.4       &      83.0	 &        50.9 &     0.0285  &  WN   &   943.4   &        16.2 &     0.0110\\
    BO s25A500B0.2        &     198.0	 &       173.3 &     0.0310  &  WN   &   976.4   &       178.8 &     0.0026\\
    BO s20A1000B0.5       &      63.3	 &        40.4 &     0.0363  &  WN   &   912.7   &       203.9 &     0.0144\\
    BO s20A500B0.1        &     394.1	 &       271.2 &     0.0010  &  WN   &   1020.0  &       375.3 &     0.0037\\
    ZM (New.) A1B1G3	  &     550.3	 &       311.4 &     0.0007  &  WN   &   1049.1  &        31.9 &     0.0393\\
    ZM (New.) A2B5G4	  &      74.0	 &        70.0 &     0.0067  &  WN   &   1068.9  &       455.2 &     0.0338\\
    ZM (New.) A4B2G5	  &     395.2	 &       386.6 &     0.0011  &  WN   &   1062.5  &       577.2 &     0.0330\\
    ZM (New.) A2B5G5	  &      55.2	 &        42.3 &     0.0031  &  WN   &   1299.8  &         8.5 &     0.0181\\
    ZM (New.) A3B4G3	  &     112.4	 &        59.7 &     0.0038  &  WN   &   1300.0  &        32.8 &     0.0237\\
    ZM (New.) A3B4G4	  &     147.3	 &       137.0 &     0.0044  &  WN   &   1316.1  &       105.3 &     0.0020\\
    ZM (New.) A2B2G3	  &     588.3	 &       165.7 &     0.0009  &  WN   &   1276.6  &       353.4 &     0.0247\\
    ZM (New.) A3B4G5	  &     102.9	 &        77.3 &     0.0032  &  WN   &   1498.0  &        25.1 &     0.0051\\
    ZM (New.) A3B2G1	  &     193.0	 &       156.0 &     0.0218  &  WN   &   1586.1  &       439.9 &     0.0057\\
    ZM (New.) A4B1G1	  &     399.3	 &       303.1 &     0.0132  &  WN   &   1559.6  &       354.4 &     0.0025\\
    ZM (New.) A1B3G1	  &     110.1	 &       104.3 &     0.0272  &  WN   &   1494.2  &       716.4 &     0.0303\\
    ZM (New.) A2B3G2	  &     218.0	 &       143.0 &     0.0026  &  WN   &   1704.2  &        22.0 &     0.0054\\
    ZM (New.) A3B2G2	  &     370.3	 &       230.8 &     0.0012  &  WN   &   1703.7  &        34.2 &     0.0150\\
    ZM (New.) A3B2G4	  &     678.6	 &       275.2 &     0.0007  &  WN   &   1723.2  &       309.8 &     0.0141\\
    ZM (New.) A4B1G4	  &     535.0	 &       191.4 &     0.0013  &  WN   &   1718.8  &       455.2 &     0.0173\\
    ZM (New.) A3B2G5	  &     318.7	 &       297.9 &     0.0022  &  WN   &   1704.4  &       649.7 &     0.0165\\
    ZM (New.) A4B1G5	  &     473.3	 &       483.3 &     0.0018  &  WN   &   1704.4  &       642.5 &     0.0320\\
    ZM (New.) A3B3G2	  &     163.9	 &       106.9 &     0.0034  &  WN   &   1699.2  &        45.9 &     0.0244\\
    ZM (New.) A4B2G2	  &     248.9	 &       171.6 &     0.0023  &  WN   &   1991.7  &       164.3 &     0.0180\\
\hline
\end{tabular}

\caption{\label{tab:parameter_rob}Parameters of the 100 waveforms used to test the robustness of the different burst pipelines: 50 waveforms are simulated core collapse, the 50 other (WN) are filtered white noise bursts. DFM waveforms (Newtonian or relativistic simulation) are taken from \cite{ref:dimmelmeier02}, BO waveforms (Newtonian) are taken from \cite{ref:ott04} and ZM (Newtonian) waveforms are extracted from \cite{ref:zwerger}. The peak frequency, bandwidth and duration of the signals have been computed as explained in Section \ref{sec:parameter}.}
\end{table}

\end{document}